\documentclass[aps,pra,showpacs,amsmath,amssymb,amsfonts,lengthcheck,twocolumn,longbibliography]{revtex4-1}

\usepackage{dsfont}
\usepackage{graphicx} 
\usepackage{graphics}

\usepackage{mathtools}
\usepackage{dcolumn}    
\usepackage{bm} 
\usepackage{graphicx}
\usepackage{amsmath}    
\usepackage{latexsym}
\usepackage{amsfonts}   
\usepackage{amssymb}
\usepackage{array}      
\usepackage{epsfig}
\usepackage{txfonts}
\usepackage{xcolor}
\usepackage[colorlinks=true,linkcolor=blue,urlcolor=blue,citecolor=blue,pdfusetitle]{hyperref}
\usepackage{hyperref}
\usepackage{ulem}

\newcommand {\fabs}[1] {\left| #1 \right|}
\newcommand{\norm}[1]{\left\lVert#1\right\rVert}

\newcommand{\ket}[1]{\left|#1\right\rangle}
\newcommand{\bra}[1]{\langle#1|}
\newcommand{\braket}[2]{\langle#1|#2\rangle}
\newcommand{\ketbra}[2]{|#1\rangle\langle#2|}

\usepackage[none]{hyphenat}

\newcommand{\bla}{bla\\bla\\bla\\bla\\bla}
\begin{document}

\title{Counterdiabatic control in the impulse regime}
\author{Eoin Carolan}
\author{Anthony Kiely}
\author{Steve Campbell}
\affiliation{School of Physics, University College Dublin, Belfield, Dublin 4, Ireland}
\affiliation{Centre for Quantum Engineering, Science, and Technology, University College Dublin, Belfield, Dublin 4, Ireland}
\begin{abstract}
Coherent control of complex many-body systems is critical to the development of useful quantum devices. Fast perfect state transfer can be exactly achieved through additional counterdiabatic fields. We show that the additional energetic overhead associated with implementing counterdiabatic driving can be reduced while still maintaining high target state fidelities. This is achieved by implementing control fields only during the impulse regime, as identified by the Kibble-Zurek mechanism. We demonstrate that this strategy successfully suppresses most of the defects that would be generated due to the finite driving time for two paradigmatic settings: the Landau-Zener model and the Ising model. For the latter case, we also investigate the performance of our impulse control scheme when restricted to more experimentally realistic local control fields.
\end{abstract}
\date{\today}
\maketitle

\section{Introduction \label{intro}}
To fully exploit the promises of quantum devices~\cite{DeutschPRXQ, Preskill2018}, efficient and effective techniques to achieve coherent control are crucial. Adiabatic methods are inherently stable but can require long timescales, particularly for many-body systems. Numerous alternative approaches have been developed that can be broadly bisected into: (i) optimal control techniques~\cite{OCreview}, which efficiently find bespoke controls for a given task, often numerically, and (ii) shortcuts-to-adiabaticity~\cite{Torrontegui2013,Guery-Odelin2019} which reproduce the same high fidelity as adiabatic passage but in significantly shorter times and are often analytic in nature. Recently, hybrid approaches have been shown to be highly effective~\cite{Saberi2014, Campbell2015, KielyNJP2021, WhittyPRR, SelsE3909, SenguptaPRB}.

Counterdiabatic driving~\cite{Demirplak2003,Berry2009} is a particularly simple and effective shortcut-to-adiabaticity, achieving perfect control by adding auxiliary terms to a given system's Hamiltonian. Such an additional control term heuristically implies an overall increase in resources needed to evolve the system. Various cost measures have been developed~\cite{Zheng2016, MugaNJP2018, Muga2017PRA,Abah2017,Cakmak2019,delCampoPRL2017,Chen2012} to characterise this. These measures have been shown to be closely related to quantum speed limits~\cite{Campbell2017, delCampoPRL2017, Puebla2020} and relevant for other
control techniques ~\cite{Abah2019, LatunePRA, Santos2015SciRep, DeffnerEPL}. In the case of critical systems with vanishing energy gaps, these cost measures indicate that the energetic resources needed to implement high fidelity control diverge. Nevertheless, such systems offer significant promise in, for example, critical metrology~\cite{CriticalMetPRX}, quantum annealers~\cite{GardasSciRep}, and adiabatic quantum computing \cite{Hartmann2019,Hegade2021}. Developing techniques which reduce the resource intensiveness while still achieving high fidelity control for critical systems is therefore timely for next-generation quantum technologies.

Topological defects were shown to be inevitable for field theories as a result of cosmic phase transitions~\cite{Kibble1976}. Remarkably, it was established that similar defect formation should occur in all phase transitions traversed in a finite time and it is precisely the critical slowing down in the vicinity of a phase transition that characterises the non-equilibrium dynamics of the system in terms of the equilibrium  critical exponents~\cite{Zurek1985}. Now, the celebrated Kibble-Zurek mechanism (KZM) has been applied in a great diversity of settings~\cite{DelCampo2013, DeChiara2013, DelCampo2010, DelCampoNJP2011, Sadhukhan2020, DeffnerPRE2017, AdCPRL2019}. It predicts that the overall driven dynamics is split into two separate regimes. The evolution is ``adiabatic" where the energy gap remains sufficiently large and the system can be driven without significant excitations being created; and ``impulsive" when the system's response freezes-out and defects rapidly form.

We exploit the insight provided by the KZM to devise an efficient strategy for achieving high fidelity control. We limit the application of the counterdiabatic control term to the duration of the system’s impulse regime, achieving significant energy savings without drastically sacrificing efficacy. While the system does generate some intermediate defects during the uncontrolled evolution in the adiabatic regimes~\cite{MolmerPRA}, these regions are precisely those in which the KZM predicts that the system is able to relax. By restricting the application of counterdiabatic control to the impulse regime, we are still able to benefit from the good performance of adiabatic passage while simultaneously reducing the resource overheads compared with full evolution control.

\section{Preliminaries \label{prelim}}
\subsection{Kibble-Zurek mechanism \label{KZM}}  
The KZM provides a framework to identify when a system crosses from the adiabatic to impulse regime and vice-versa. Consider a system Hamiltonian $H(g)$ with a critical point at $g\!=\!g_c$, and a linearly varying external field $g(t)\!=\!g_0+g_d t/\tau_Q$ where $\tau_Q$ is the quench time of the system. The transition times, $t_{\mp}$, are defined as when the relative rate of parameter change is comparable to the relaxation time. Formally, they are the solutions of 
\begin{equation}
\tau(t_{\mp})=\fabs{\frac{g(t_{\mp})-g_c}{\Dot{g}(t_{\mp})}}, \label{eq_times}
\end{equation}
where $\tau\!=\!\hbar/\gamma$ is the relaxation time and $\gamma$ is the relevant energy gap~\cite{Damski2005}. For convenience we assume a symmetric ramp, taking $g(\tau_Q/2)\!=\!g_c$ which fixes $g_d\!=\!2(g_c-g_0)$. This simplifies Eq.~\eqref{eq_times} to $\tau(t_{\mp})=\fabs{t_{\mp}-\frac{\tau_Q}{2}}$. As the system nears the critical point the correlation length, $\xi$, and relaxation time diverge. Renormalization-group theory~\cite{Sachdev, FisherRMP} gives the scaling as $\xi \! \sim \! \xi_0 \fabs{g-g_c}^{-\nu}$ and $\tau \! \sim \! \tau_0 \fabs{g-g_c}^{-z \nu}$ where $\nu$ is the spatial exponent and $z$ is the dynamical critical exponent. Kibble-Zurek scaling then predicts that the adiabatic-impulse regime crossover times obey
\begin{eqnarray}
t_{\mp}=\frac{\tau_Q}{2}\mp \tau_0^{\frac{1}{1+ z \nu}}\left(\frac{\tau_Q}{2 \fabs{g_0-g_c}}\right)^{\frac{z \nu}{1+ z \nu}}, \label{imp_scaling}
\end{eqnarray}
and the correlation length at this time is
\begin{eqnarray}
\xi = \xi_0 \left(\frac{2 \tau_0 \fabs{g_0-g_c}}{\tau_Q}\right)^{\frac{-\nu}{1+z \nu}}.
\end{eqnarray}
The density of defects for a system of dimension $d$ then scales as $\xi^{-d} \sim \tau_Q^{-\frac{\nu d}{1+z \nu}}$. Note that in some systems the relevant energy gap $\gamma$ is known explicitly, e.g. in the Landau-Zener model considered in Sec.~\ref{LZM}, and thus the exact expression for the relaxation time and resulting impulse regime can be employed, while in the case of genuine many-body settings, such as the transverse field Ising model considered in Sec.~\ref{TFIM}, the gap can be approximated.

\subsection{Counterdiabatic driving \label{TCC}}
Consider a Hamiltonian with spectral decomposition
\begin{equation}
    H_0(t)=\sum_n \epsilon_n(t)\ket{\phi_n(t)}\bra{\phi_n(t)},
\end{equation}
with $\epsilon_n(t)$ and $\ket{\phi_n(t)}$ the instantaneous energy eigenvalues and eigenstates, respectively, and we wish to evolve the system along an instantaneous energy eigenstate, e.g. the ground state. For long operation times the adiabatic approximation predicts
\begin{equation}\label{eq:rrr}
    \ket{\psi_n(t)}\approx\text{exp}[\alpha_n(t)+\beta_n(t)]\ket{\phi_n(t)},
\end{equation}
where $\alpha_n=-\frac{1}\hbar\int_0^t \epsilon_n(s)ds$ and $\beta_n= i\int_0^t\braket{\psi_n(s)}{\partial_{s}\phi_n(s)}ds$ are the dynamic and geometric phases and the initial condition is $\ket{\psi_n(0)}=\ket{\phi_n(0)}$.

Remarkably, precisely the same adiabatic evolution captured by Eq.~\eqref{eq:rrr} can be achieved exactly on arbitrarily short timescales by using additional control fields defined by the Hamiltonian~\cite{Demirplak2003, Berry2009}
\begin{equation}
\label{eq:cdcd}
    H_{CD}(t)=i\hbar\sum_n\big[ 
     \ketbra{\partial_t\phi_n(t)}{\phi_n(t)}
    -\braket{\phi_n(t)}{\partial_t\phi_n(t)}\ketbra{\phi_n(t)}{\phi_n(t)}\big].
\end{equation}
Evolving the system using the total Hamiltonian $H_0+H_{CD}$ therefore forbids any nonadiabatic transitions.

Adiabatic timescales diverge in critical systems due to vanishing energy gaps in the thermodynamic limit. Employing Eq.~\eqref{eq:cdcd} for a given initial state allows one to traverse the quantum phase transition of a finite system in finite time~\cite{delCampoPRL2012, DamskiJStat, Campbell2015}. However, the magnitude and complexity of the control fields near the critical point grow significantly with system size~\cite{Demirplak2008, Zheng2016} implying that control comes at a high energetic cost~\cite{Campbell2017, delCampoPRL2017, Abah2019}.

The KZM demonstrates that spurious excitations or defects are mainly generated within the impulse regime, with the system evolving almost adiabatically otherwise. In order to minimise energetic cost we propose limiting the use of a control strategy to only during the impulse regime, as opposed to employing control for the entire evolution. To test the effectiveness of this approach, we consider the Hamiltonian 
\begin{equation}
    H_\kappa (t)=H_0(t)+\left[\delta_{1\kappa}+\delta_{2\kappa}S(t)\right]H_{CD}(t), \label{ham}
\end{equation}
where  $\kappa\! \in\! \{0,1,2\}$ corresponds to uncontrolled, fully controlled, and impulse controlled systems, respectively, and $\delta_{ij}$ is the Kroeneker delta. The control field $H_{CD}$ is smoothly turned on during the impulse regime with a switching function $S(t)=f(t-t_{-})f(t_{+}-t),$ where $f(x)=1/\left(1+e^{-m x}\right)$ is the logistic function and $m$ a constant determining the abruptness of the switch.

\begin{figure*}[t]
\begin{center}
\includegraphics[width=0.333\textwidth]{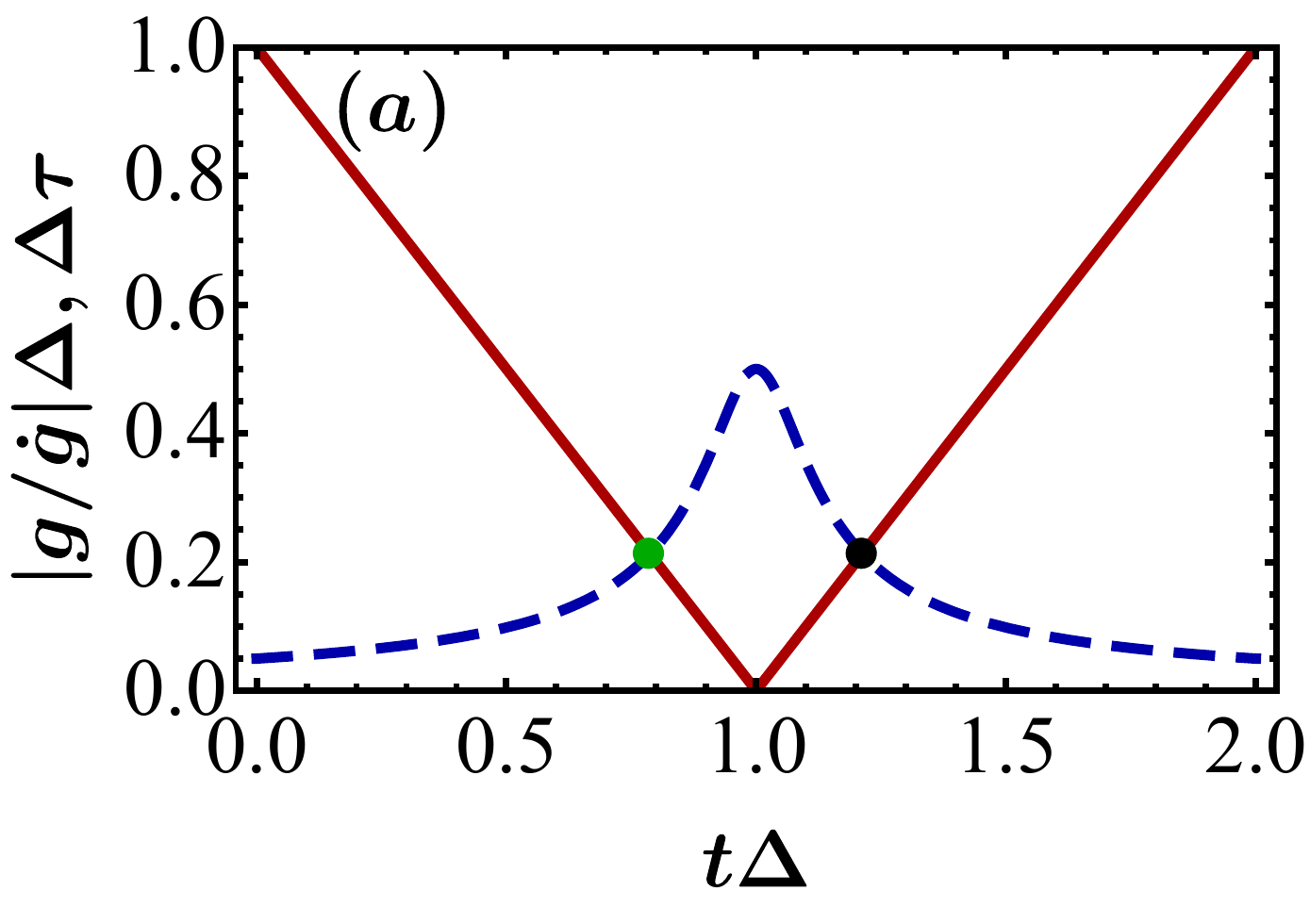}~~\includegraphics[width=0.333\textwidth]{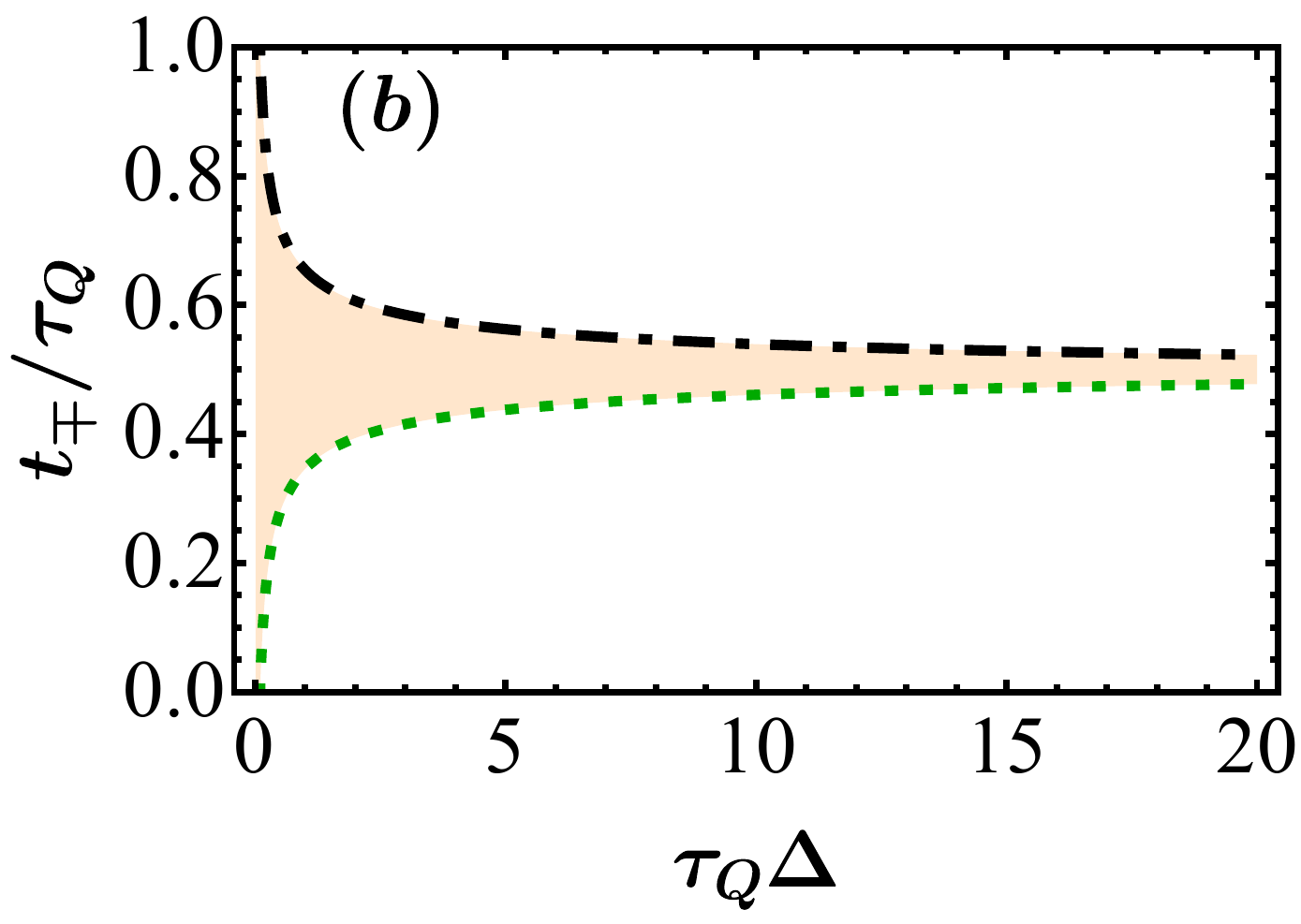}~~\includegraphics[width=0.333\textwidth]{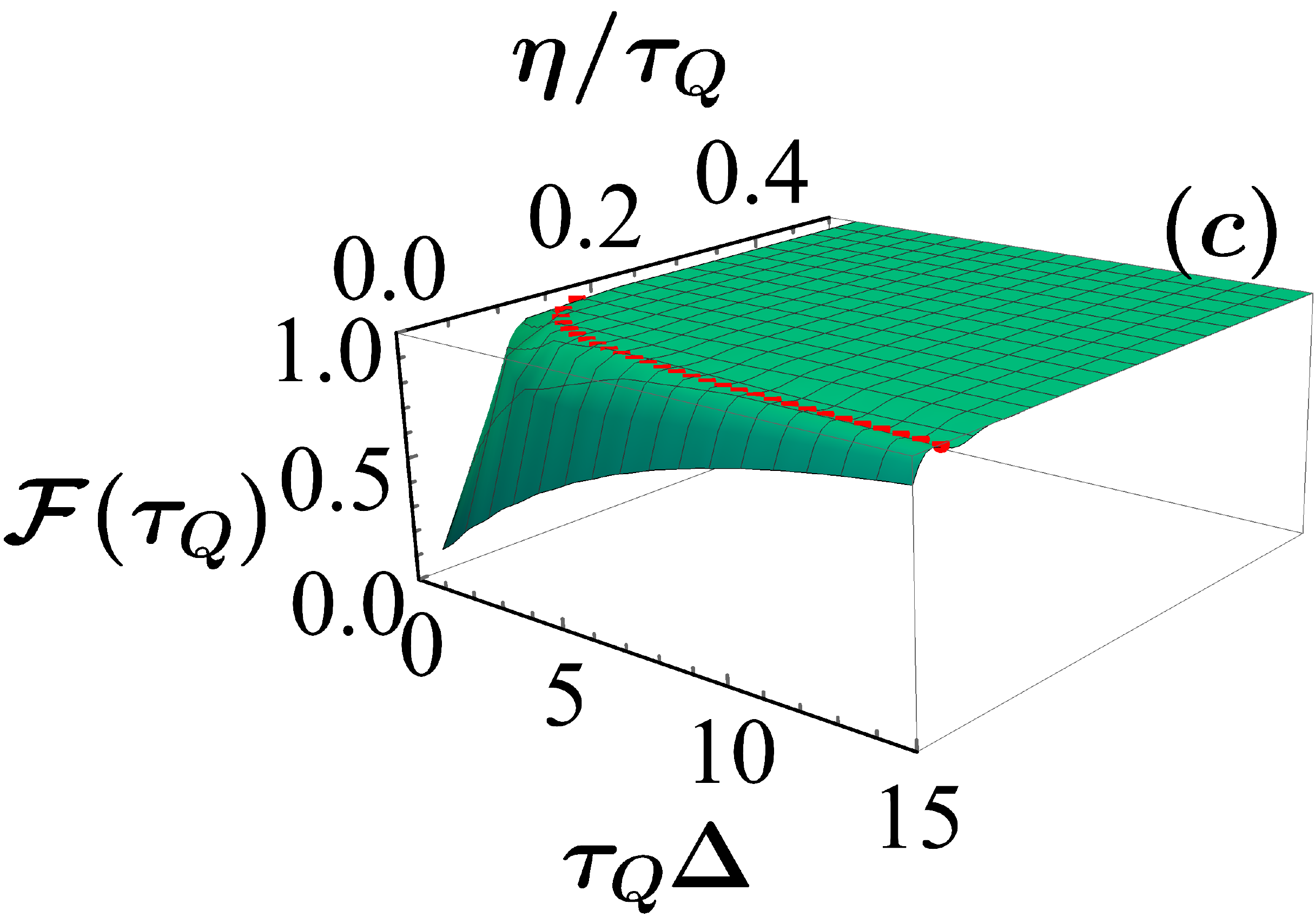}
\end{center}
\caption{Impulse regime for the Landau-Zener model: (a) Comparison between relaxation timescale $\tau$ (red solid line) and the relative rate of external parameter change $\fabs{g/\dot{g}}$ (blue dashed line) for a quench duration of $\tau_Q\Delta \!=\! 2$. Intersection gives critical times $t_{\mp}$ (green/black dots) (b) Impulse regime times $t_{\mp}$ (green dotted and black dot-dashed lines respectively)  for different quench times $\tau_Q$. (c) Final state fidelity versus quench time $\tau_Q$ and duration of counterdiabatic driving $2\eta$. Impulse regime $\eta=\mu$ (red points), $g_0=-10 \Delta$ and $m=400 \Delta^{-1}$. \label{fig_impulse}}
\end{figure*}

\subsection{Figures of merit \label{meas}}
The performance of each protocol will firstly be quantified by focusing on the fidelity of the state of the system, $\ket{\psi(t)}$, evolving according to Hamiltonian Eq.~\eqref{ham} with the instantaneous ground state, i.e.
\begin{equation}
    \mathcal{F}(t)=\fabs{\bra{\psi(t)}{\phi_0(t)}}^2,
\end{equation}
We assume $\ket{\psi(0)}\!=\!\ket{\phi_0(0)}$ as the initial condition in all cases i.e. the system starts in the ground state. The intensity of the additional control field provides a meaningful quantifier of the energetic cost of the control~\cite{Demirplak2008, Zheng2016}. We can quantify energetic cost of full counterdiabatic control ($\kappa=1$) as \footnote{In Ref.~\cite{Zheng2016} the cost is defined as $\mathcal{C}\!\!=\!\!\frac{1}{\tau_Q}\int_0^{\tau_Q}ds\norm{H_{CD}(s)}^n$, where the choice of $n$ depends on the physical implementation. Here, we take $n\!=\!1$ for simplicity and remark that qualitatively similar behaviors are exhibited for other suitable choices of $n$.}
\begin{equation}
\mathcal{C}=\frac{1}{\tau_Q}\int_0^{\tau_Q}ds\norm{H_{CD}(s)}, \label{cost}
\end{equation}
where $\norm{\cdot}$ is the Frobenius norm. It is clear that $\mathcal{C}$ scales as $\sim \hbar /\tau_Q$\cite{Zheng2016} from the form of $H_{CD}$. The relative energetic savings achieved by employing only impulse control ($\kappa=2$) is $\delta E/ \mathcal{C}$, where $\delta E$ is the absolute energetic savings
\begin{eqnarray}
 \delta E &=& \frac{1}{\tau_Q}\int_0^{\tau_Q}ds \left[1- S(s)\right]\norm{H_{CD}(s)} ,\nonumber \\
 &\approx & \frac{1}{\tau_Q} \left[\int_0^{t_-}ds\norm{H_{CD}(s)}+ \int_{t_+}^{\tau_Q}ds\norm{H_{CD}(s)}  \right]. \label{abs_saving}
\end{eqnarray}

 As the quench time becomes shorter, $\tau_Q\!\rightarrow\!0$, the impulse regime dominates $\left\{t_{-},t_{+}\right\}\!\rightarrow\! \left\{0,\tau_Q\right\}$. Clearly then for impulse control, we should expect $\delta E\!\rightarrow\! 0$ and $\mathcal{F}(\tau_Q)\!\rightarrow \! 1$ in the short quench time limit. For long quench times, $\tau_Q\!\rightarrow\! \infty$, the impulse regime vanishes $t_{\mp} \! \rightarrow \! \tau_Q/2$. Therefore the relative savings $\delta E/\mathcal{C}\rightarrow 1$ and $\mathcal{F}(\tau_Q)\rightarrow 1$ due to the adiabatic theorem.

 Note that the cost is lower bounded as 
 \begin{equation}
 \mathcal{C} \geq \frac{\hbar}{\tau_Q} \int_{g_0}^{g(\tau_Q)} W[g] dg,
 \end{equation}
 where $W[g]=\sqrt{\sum_{n,m\neq n} \left|\frac{\bra{\phi_m} \partial_g H_0 \ket{\phi_n}}{\epsilon_n-\epsilon_m}\right|^2}$. In what follows, we consider a linear ramp for simplicity since any monotonic choice of $g$ achieves the minimum of the cost measure employed.

\section{Landau-Zener Model \label{LZM}}
We begin our analysis with the Landau-Zener (LZ) model. It describes the transitions of a two-level quantum system interacting with an external field as it passes through resonance~\cite{Zener1932}. The Hamiltonian is
\begin{equation}\label{eq:lzkz}
    H_0(t)=\hbar \Delta\sigma_x+\hbar g(t)\sigma_z,
\end{equation}
where $\Delta\!>\!0$ determines the minimal energy gap at the avoided crossing. Despite not exhibiting a bonafide quantum phase transition, the LZ model captures all basic features of the KZM~\cite{Damski2005, Damski2006, DelCampo2014, Zurek2005}, including recovering the expected critical exponents: $\nu=1$ and $z=1$. The energy eigenstates are
 \begin{eqnarray}
 \ket{\phi_0(t)} &=& \cos\left[\theta(t)\right]\ket{0}+\sin\left[\theta(t)\right] \ket{1} , \\
  \ket{\phi_1(t)} &=& \sin\left[\theta(t)\right] \ket{0}- \cos\left[\theta(t)\right]\ket{1}, 
 \end{eqnarray}
where $\tan\left[\theta(t)\right]\!=\!-\left[g(t)+\sqrt{\Delta^2+g(t)^2}
\right]/\Delta$ and the energy gap between ground and excited states is $\gamma\!\!=\!\!\epsilon_1-\epsilon_0\!\!=\!\! 2\sqrt{g(t)^2+\Delta^2} \hbar$. The counterdiabatic Hamiltonian is then~\cite{Berry2009}
\begin{equation}
    H_{CD}(t)= \hbar \dot{\theta} \sigma_y = -\frac{\dot{g}(t)\Delta \hbar }{2\left[\Delta^2+g(t)^2\right]}\sigma_y.
\end{equation}

Solving for the real roots of Eq.~\eqref{eq_times} we find the adiabatic-impulse crossover times as $t_{\mp}\!=\!\tau_Q/2 \mp \mu$ where
\begin{eqnarray}
\mu= \frac{1}{2}\sqrt{\frac{\sqrt{\tau_Q^4\Delta^4 +4 g_0^2 \tau_Q^2}-\tau_Q^2\Delta^2 } {2 g_0^2}},
\end{eqnarray}
as shown in Fig.~\ref{fig_impulse}(a) for a representative quench time of $\tau_Q\Delta\!=\!2$. Note that the impulse regime has a duration $2\mu$ and $0\! \leq\! \mu\! \leq \! \tau_Q/2$. For short operation times, the behaviour is $\mu \approx \sqrt{\tau_Q/\fabs{g_0}}/2$ which matches KZM scaling predictions for the  $\nu\!=\!z\!=\!1$ universality class.Fig.~\ref{fig_impulse}(b) shows the impulse regime for a linear ramp of fixed magnitude for various quench durations. It highlights that slow ramps recover effectively adiabatic dynamics with the width of the impulse regime closing as $\tau_Q$ grows, while for fast ramps the system is effectively always in the impulse regime. 

In Fig.~\ref{fig_impulse}(c) we verify that control in the impulse regime is crucial for achieving a high fidelity final state. We implement a protocol in which the counterdiabatic control field is switched on for a duration of $\eta$ before and after the system reaches the avoided crossing, i.e. $H_{CD}$ is switched on/off at
\begin{equation}
    \tilde{t}_{\mp}=\frac{\tau_Q}{2}\mp\eta,
\end{equation}
where $\eta\!\in\![0,\tau_Q/2]$. This smoothly interpolates between the three cases captured by Eq.~\eqref{ham} with $\eta=\left\{0,\tau_Q/2,\mu\right\}$ corresponding to $\kappa=\left\{0,1,2\right\}$ respectively. The red-dashed lined delineates the adiabatic-impulse crossover and we see that there is a precipitous drop when the control is applied for durations smaller than the impulse regime, i.e. $\eta\!<\!\mu$. For protocols with $\eta\!>\!\mu$ we see that there is little gain in target fidelity by continuing to employ the counterdiabatic term.

\begin{figure*}[t]
\begin{center}
\includegraphics[width=0.25\linewidth]{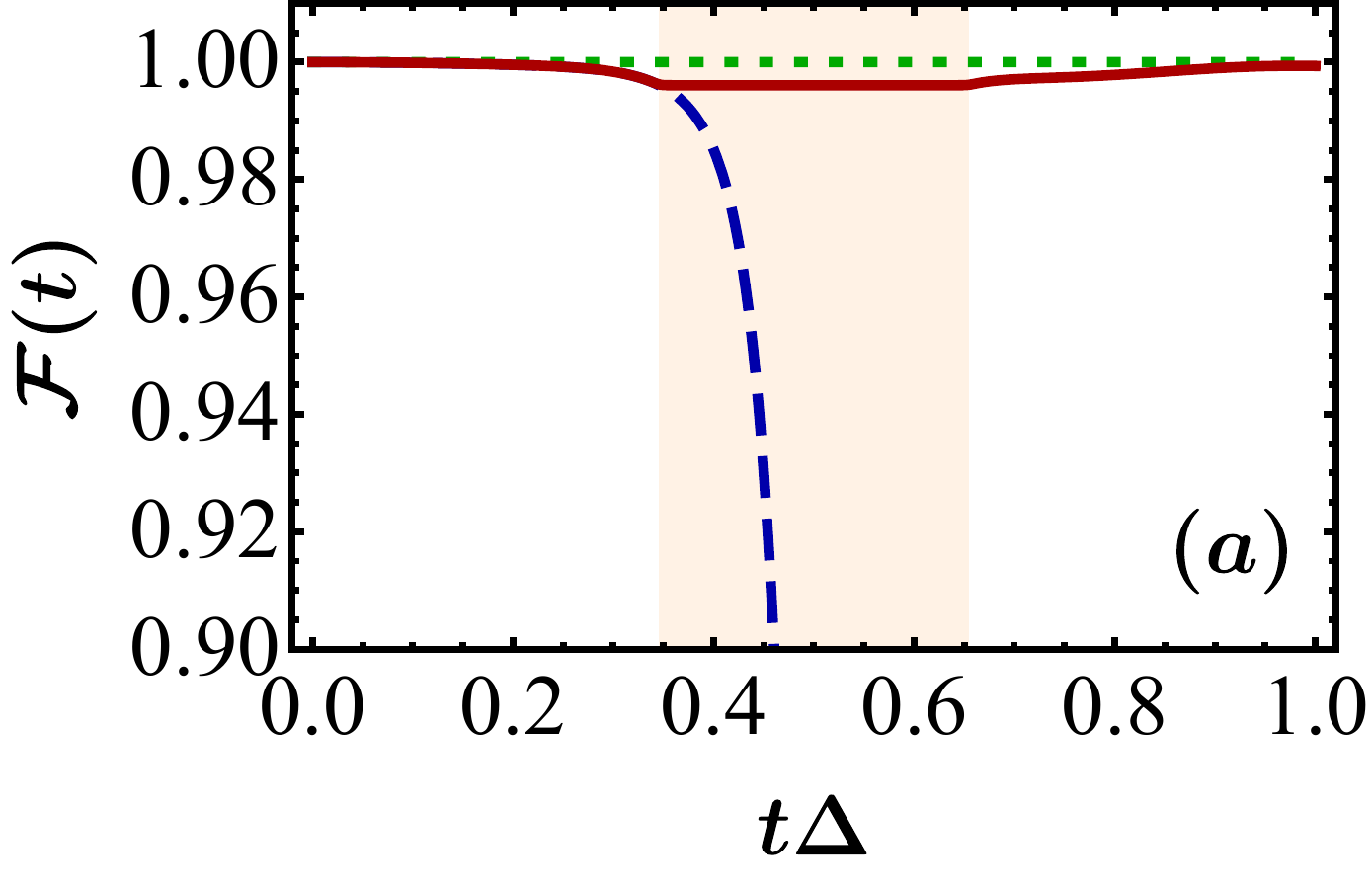}~~\includegraphics[width=0.25\linewidth]{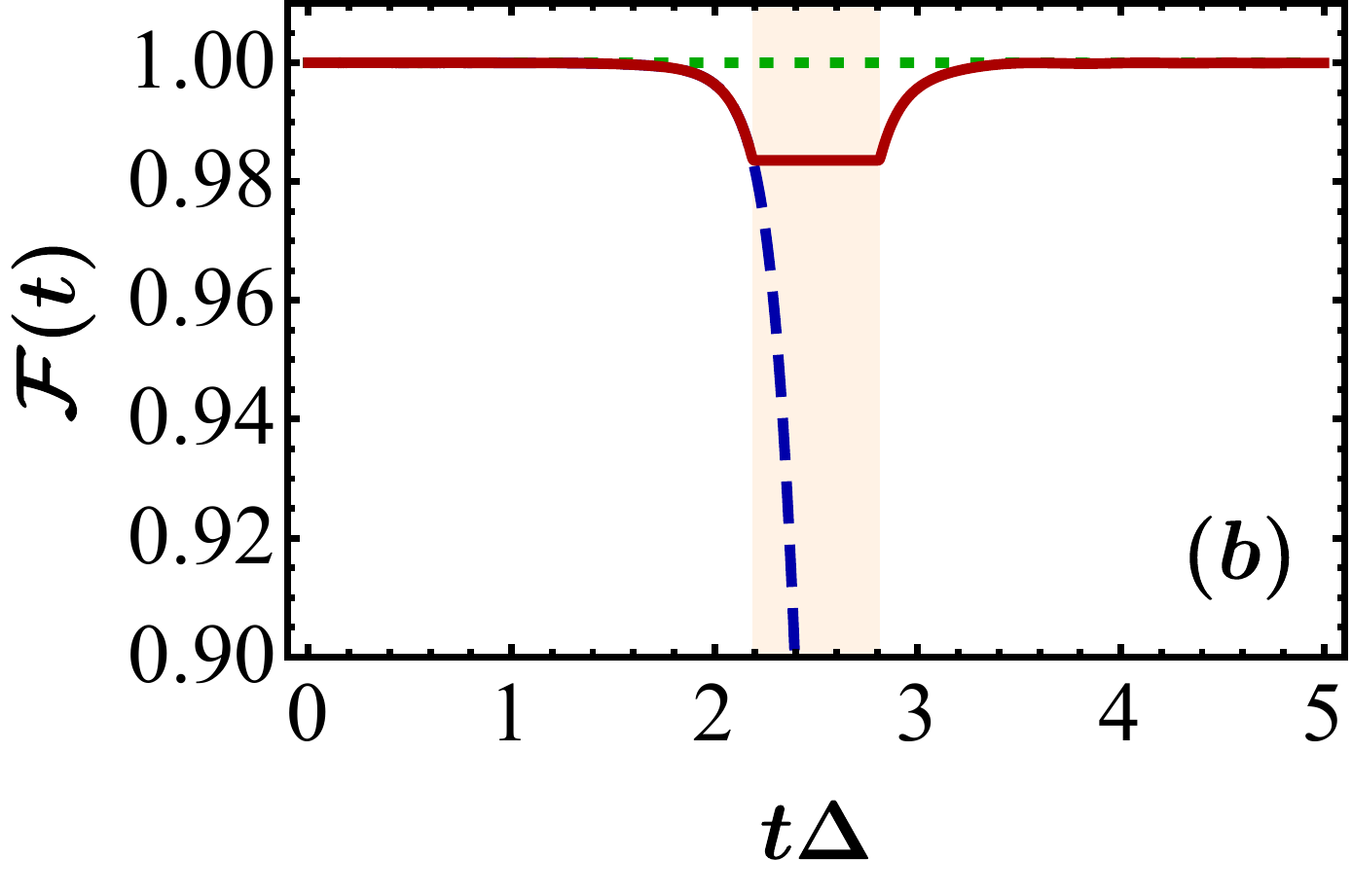}~~\includegraphics[width=0.25\linewidth]{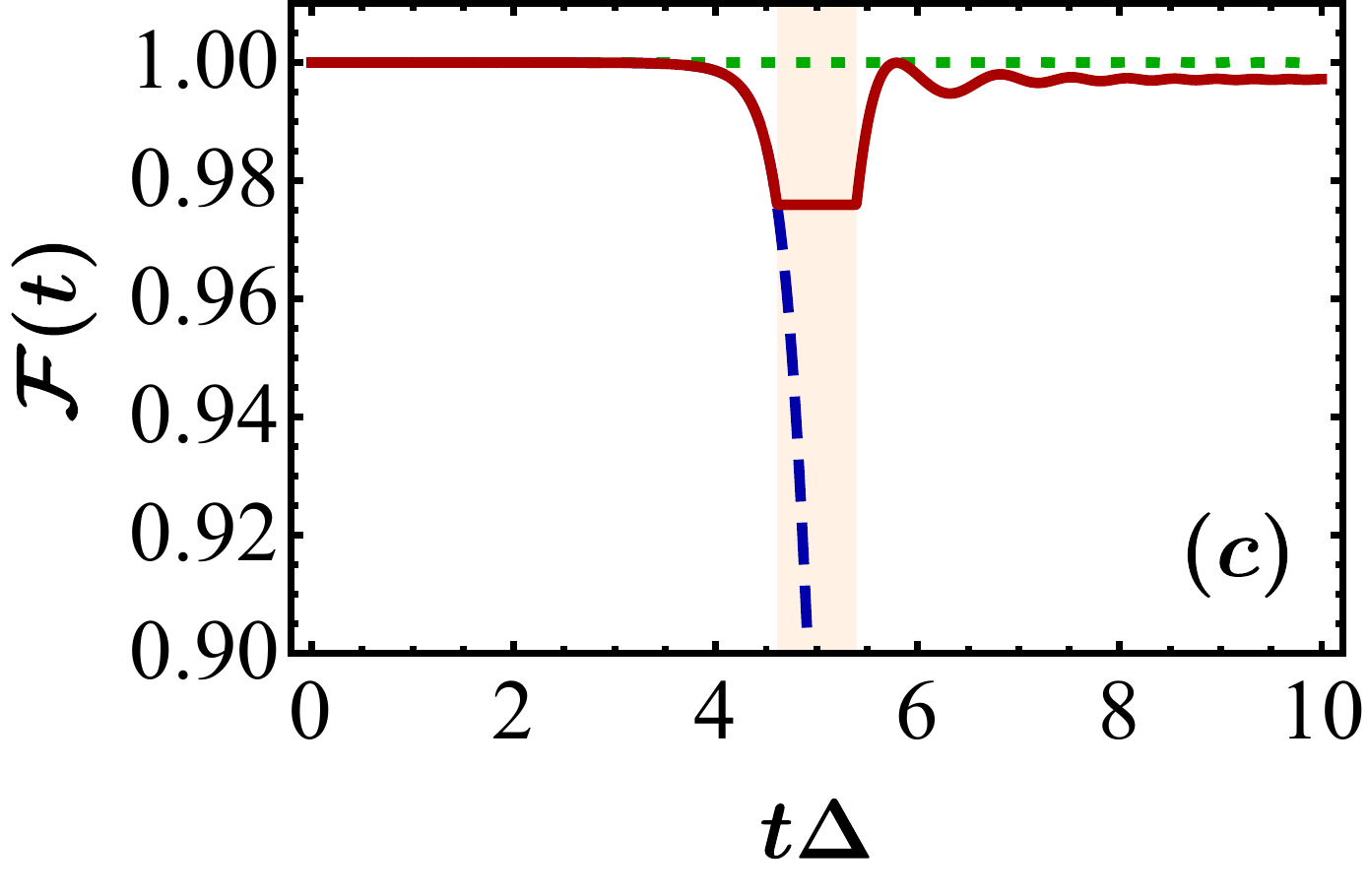}~~\includegraphics[width=0.25\linewidth]{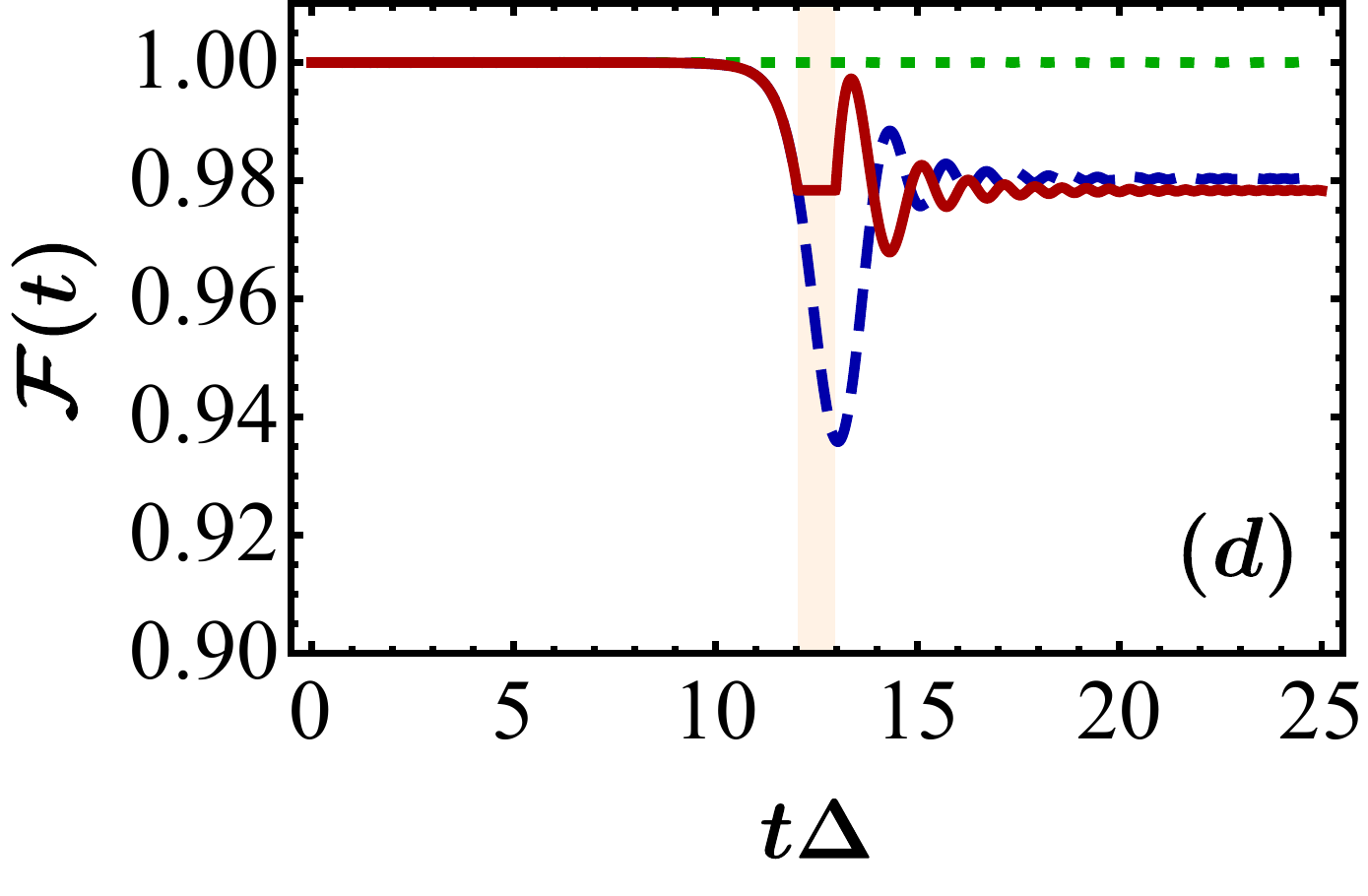}
\end{center}
\caption{Fidelity of the evolving state with the instantaneous ground state of the Landau-Zener model for no control (blue dashed line), impulse control (red solid line) and full control (green dotted line), for different quench times. Orange shaded area indicates the impulse regime $[t_{-},t_{+}]$. (a) $\tau_Q\Delta\!=\! 1$ (b) $\tau_Q\Delta \!=\! 5$ (c) $\tau_Q\Delta\!=\!10$ (d) $\tau_Q\Delta\!=\!25$. Other parameters: $g_0\!=\!-10\Delta$ and $m\!=\!400 \Delta^{-1}$}
 \label{fig_runs}
\end{figure*}

We now take a more systematic look at the three protocols, having established that high-fidelity final states are principally reliant on implementing control when the system is in the impulse regime. The instantaneous fidelity with the ground state for the three cases are shown in Fig. \ref{fig_runs} for various quench durations. The evolution under full counterdiabatic control remains in the ground state at all times by construction and therefore results in a perfect fidelity. If no control is applied the system maintains a high instantaneous fidelity initially, but this rapidly decreases once it enters the impulse regime, delineated by the orange shaded area. For short quench times, once the fidelity drops off there is little revival. However, for sufficiently long times, where the impulse regime is short enough that significant defects are not generated (e.g. $\tau_Q\!=\!25 \Delta^{-1}$ in Fig. \ref{fig_runs}(d) [dashed, blue curve]), after an initial dip the fidelity increases again outside impulse regime. This dip and revival behaviour is a generic feature of adiabatic passage and is a result of the adiabatic error on the instantaneous fidelity scaling as $1/\tau_Q$, while the error on the final state fidelity scales as $1/\tau_Q^2$~\cite{MolmerPRA}. 

For impulse control (solid, red curves) the instantaneous fidelity initially follows the uncontrolled case. However, when entering the impulse regime the counterdiabatic control is switched on which negates any non-adiabatic transitions between the energy eigenstates and therefore freezes the instantaneous fidelity in this region. By freezing the system only in the impulse regime we are able to suppress most of the defects from forming such that the resulting final free evolution often leads to excellent state transfer. The resulting final fidelities are comparable to the case of full control despite the control field only being on for a fraction of the total quench time. As we increase $\tau_Q$, resulting in a closing of the impulse regime, our impulse control scheme no longer provides an advantage over the uncontrolled evolution. Upon exiting the impulse regime the dynamics is adiabatic, leading to approximately constant fidelity, as seen in Fig. \ref{fig_runs}(d). Therefore any population lost in the first stage cannot be recovered. Note that we have focused on symmetric ramps for simplicity, but the strategy of impulse control can be easily generalised to asymmetric ramps. 

\begin{figure}[t]
\begin{center}
\includegraphics[width=0.82\linewidth]{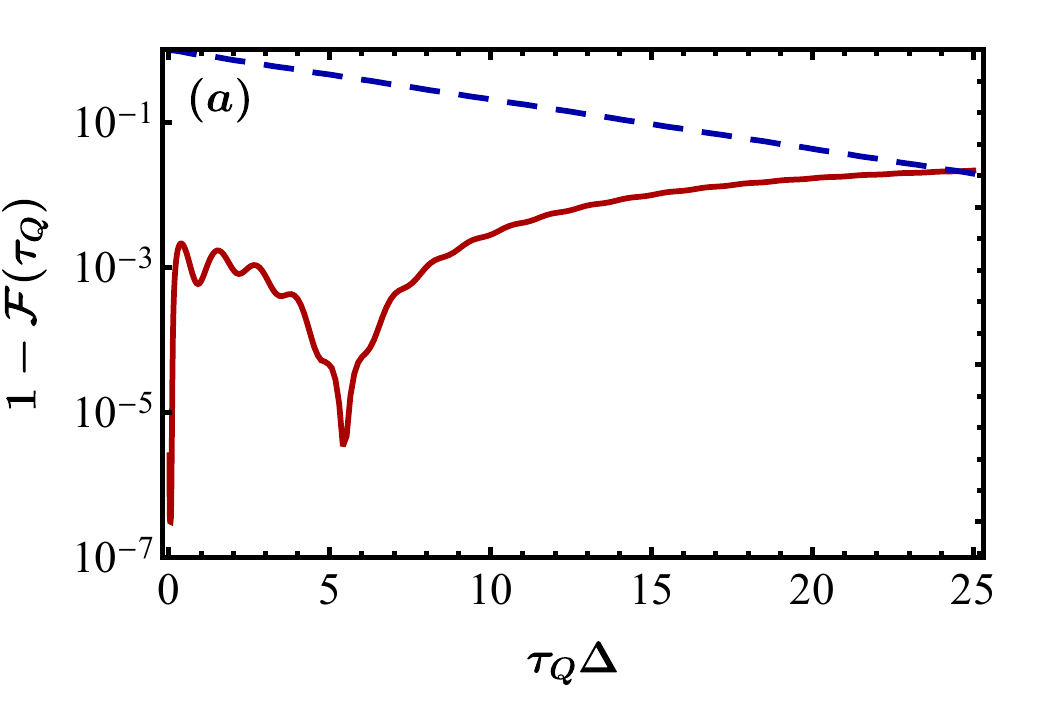} \\ \includegraphics[width=0.82\linewidth]{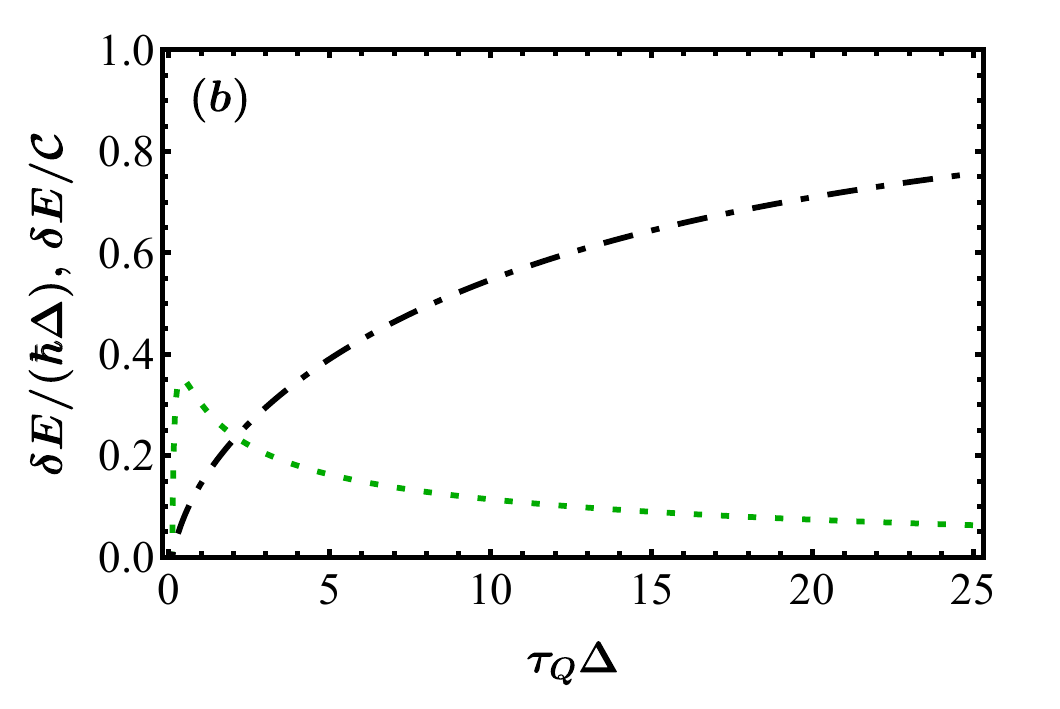}
\end{center}
\caption{Comparison between final state infidelity using impulse control and the resulting energetic savings. (a) Infidelity at the end of the process using impulse control (red solid line). Also shown is the result after no control (blue dashed line). (b) Difference in energetic cost $\delta E$ (green dotted line) and relative difference in energetic cost $\delta E/\mathcal{C}$ (black dot-dashed line). Other parameters: $g_0\!=\!-10\Delta$ and $m\!=\!400 \Delta^{-1}$.}
\label{fig_LZM_compare}
\end{figure}

Turning our attention to the energetic savings, Fig.~\ref{fig_LZM_compare} demonstrates that significantly better efficiency can be achieved with only a small loss in final state fidelity. In panel (a) we show the final state infidelity, i.e. $1-\mathcal{F}$, achieved for impulse control as a function of total quench duration and for reference we also show the no-control case which follows the well known Landau-Zener formula $\exp\left(-\pi \Delta^2/ \fabs{\dot{g}} \right)$~\cite{Zener1932, Vitanov1996}. For large quench durations (corresponding to a small impulse regimes) $\tau_Q\Delta\!\sim\!25$, implementing control turns out to be detrimental. For small quench times, where the impulse regime dominates most of the protocol, the final infidelities are vanishingly small since this case overlaps significantly with the full control case. As $\tau_Q$ is increased we see a small increase in the infidelity, which nevertheless remains $\lesssim0.001$, indicating that the protocol is still highly effective. Impulse control is shown to be particularly effective around $\tau_Q\Delta\!=\!5$ for the chosen final target state. Panel (b) demonstrates that while maintaining a high level of efficacy, impulse control allows for a significant reduction in the energetic cost, achieving infidelities $\sim\!\!10^{-5}$ while making a relative energetic saving of $\sim\!\!40\%$. The absolute energetic saving clearly tends to zero in the short and long quench time limit and the relative energetic savings tends to $1$ in the long quench time limit, all of which agrees the previous analytical predictions, see Appendix~\ref{Cost_app}.

\section{Transverse-Field Ising Model \label{TFIM}}

We now consider the transverse field Ising model (TFIM)
\begin{equation}
    H_0(t)=- \hbar \omega\sum_{i=1}^{N}\left[g(t) \,\sigma_i^x+\sigma_i^z\sigma_{i+1}^z\right].
\end{equation}
We impose periodic boundary conditions $\sigma_{N+1}^{x,y,z}\!=\!\sigma_1^{x,y,z}$ and $N$ even. The TFIM is in the same universality class as the LZ model, exhibiting a second-order quantum phase transition at $g_c\!\!=\!\!1$~\cite{Zurek2005}. To find the counterdiabatic term and impulse regime we use the Jordan-Wigner transformation to map the model to a non-interacting fermion basis, and Fourier transform to decouple the system into $N/2$ LZ type settings in momentum space, see Appendix~\ref{TFIM_app} for details. For each $k$ subspace the counterdiabatic Hamiltonian has been exactly determined~\cite{delCampoPRL2012,DamskiJStat}
\begin{eqnarray}
    H_{CD,k}&=& \frac{\hbar \dot{g} \sin(kb)}{2\left[g^2-2 g \cos(kb)+1\right]}
    \sigma_k^y, \label{Hcd_Is}
    \end{eqnarray}
where $b$ is the spacing between spins. The combined effect of these Hamiltonians can be also expressed in the original spin basis with~\cite{delCampoPRL2012, DamskiJStat}
\begin{eqnarray}
    H_{CD}&=&-\dot{g}\Big[ \sum_{m=1}^{M-1}u_m(g)H_{CD}^{[m]}+ \delta_{M,N/2}\frac{1}{2}u_{N/2}(g)H_{CD}^{[N/2]}     \Big], \label{eq:trunc0}\\ 
    H_{CD}^{[m]}&=&\sum_{n=1}^N\Bigg[\sigma_{n}^x\Big(\prod_{j=n+1}^{n+m-1}\sigma_j^{z}\Big)\sigma_{n+m}^y+\sigma_{n}^y\Big(\prod_{j=n+1}^{n+m-1}\sigma_j^{z}\Big)\sigma_{n+m}^x\Bigg],\\ 
u_m(g)&=&\frac{g^{2m}+g^{N}}{8g^{m+1}(1+g^N)}\label{eq:trunc3}.
\end{eqnarray}
Here $M$ denotes the maximum range of the interactions, with the exact counterdiabatic term given by $M\!=\!N/2$. This Hamiltonian is clearly highly non-local incurring high complexity and energetic costs~\cite{delCampoPRL2017}. We will later see the efficiency of truncating the maximum range of interactions included in $H_{CD}$ by reducing $M$~\cite{DamskiJStat}.

The energy gap for each momentum subspace is given by $\gamma_k\!=\!4 \hbar \omega \sqrt{g(t)^2-2g(t)\cos(kb)+1}$ which vanishes in the thermodynamic limit at the critical point. For a finite number of spins the gap between ground and first excited state remains finite, shrinking as $\sim\! 1/N$, and only the lowest subspace, $k_0$, is critical. To determine adiabatic-impulse crossover times we approximate this gap as $\gamma_0\!\approx\! 4 \hbar \omega \fabs{g(t)-1} $~\cite{Zurek2005, DziarmagaPRL}. The resulting crossover times, assuming $g_0\!<\!1$, are again found by solving for the real roots of Eq.~\eqref{eq_times} giving
\begin{eqnarray}
t_{\mp}=\frac{\tau_Q}{2} \mp \sqrt{\frac{\tau_Q}{8 \omega (1-g_0)}},
\end{eqnarray}
which agrees with the predicted KZM scaling, cfr Eq.~\eqref{imp_scaling}. Note that the impulse regime vanishes for long quench times $(t_+-t_-)/\tau_Q \rightarrow 0$ but does not behave correctly for short quench times $\tau_Q < 1/(2 \omega [1-g_0])$ due to the approximation of the energy gap.

\begin{figure}[t]
\begin{center}
\includegraphics[width=0.47\linewidth]{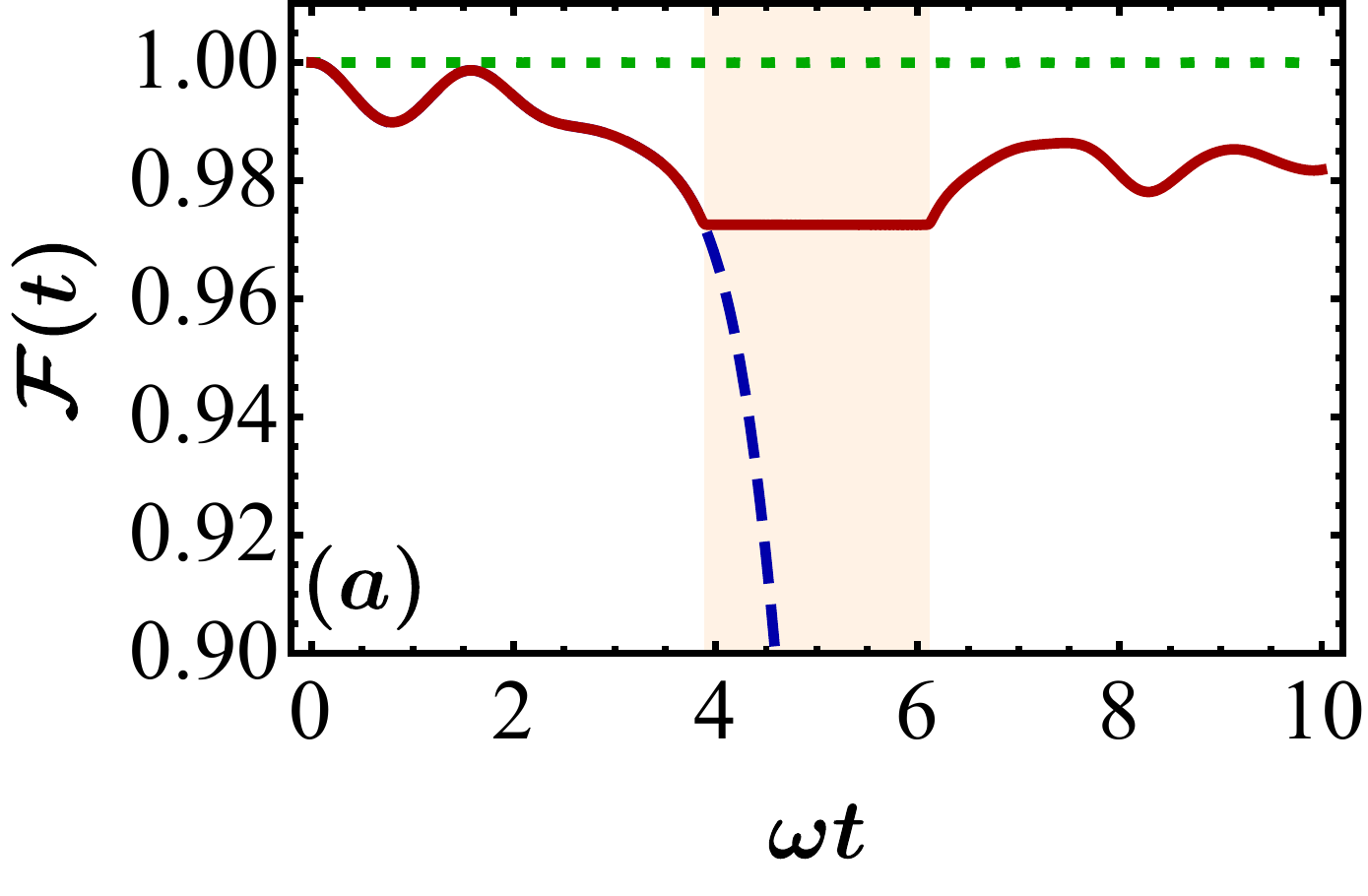}~~\includegraphics[width=0.47\linewidth]{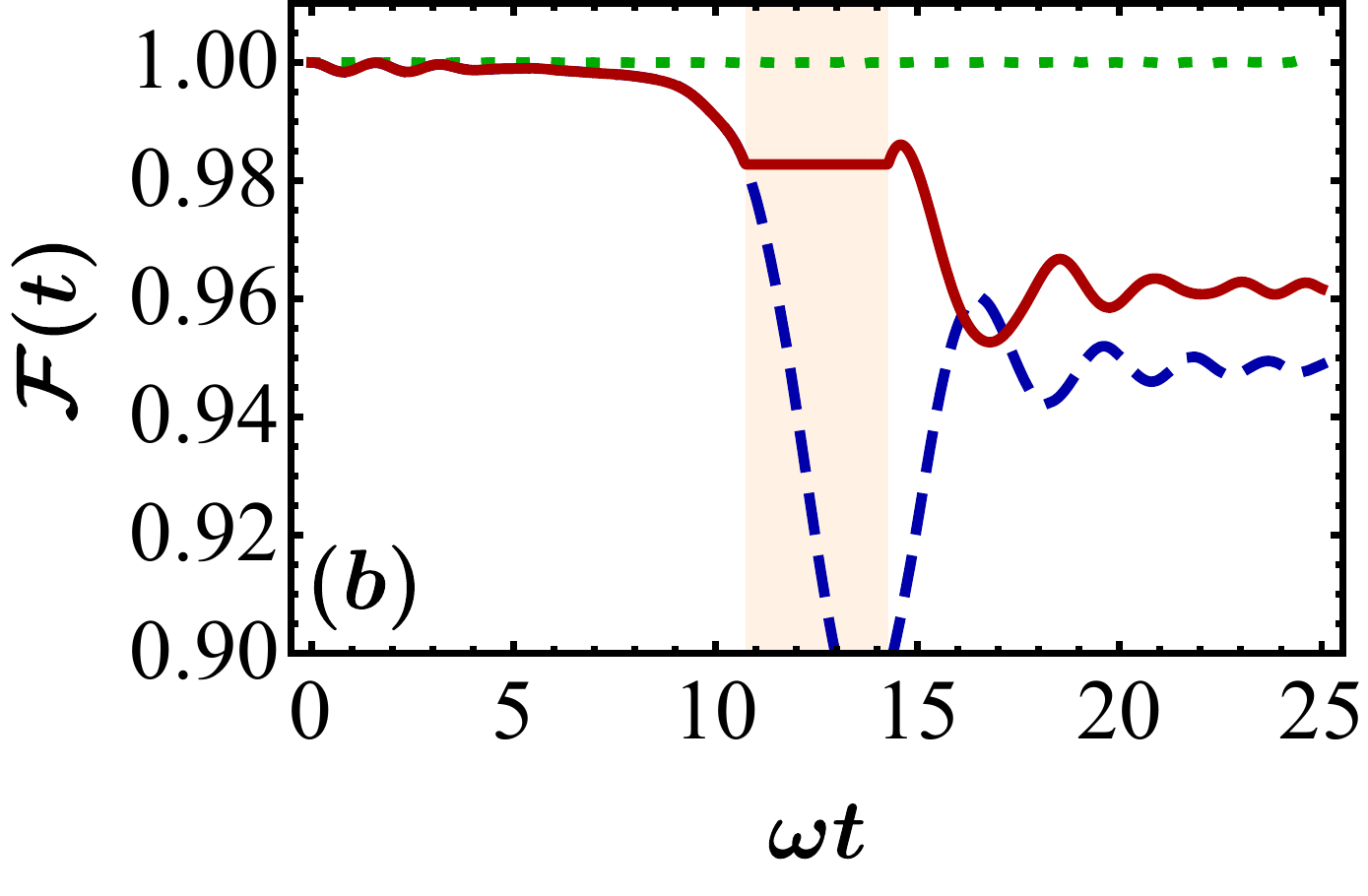}\\
\hspace*{0.4cm}\includegraphics[width=0.47\columnwidth]{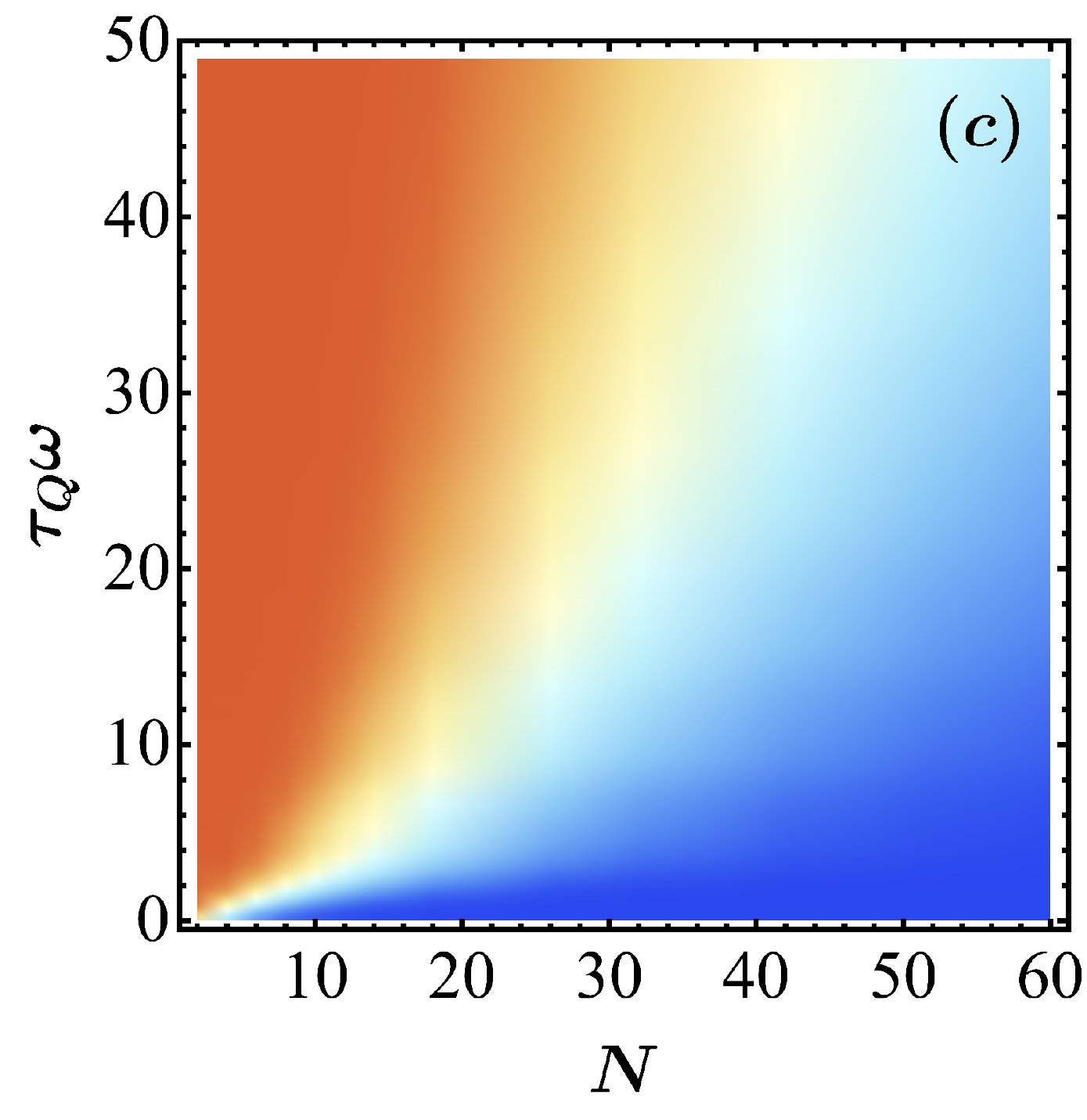}\hspace{0.1cm}\includegraphics[width=0.47\columnwidth]{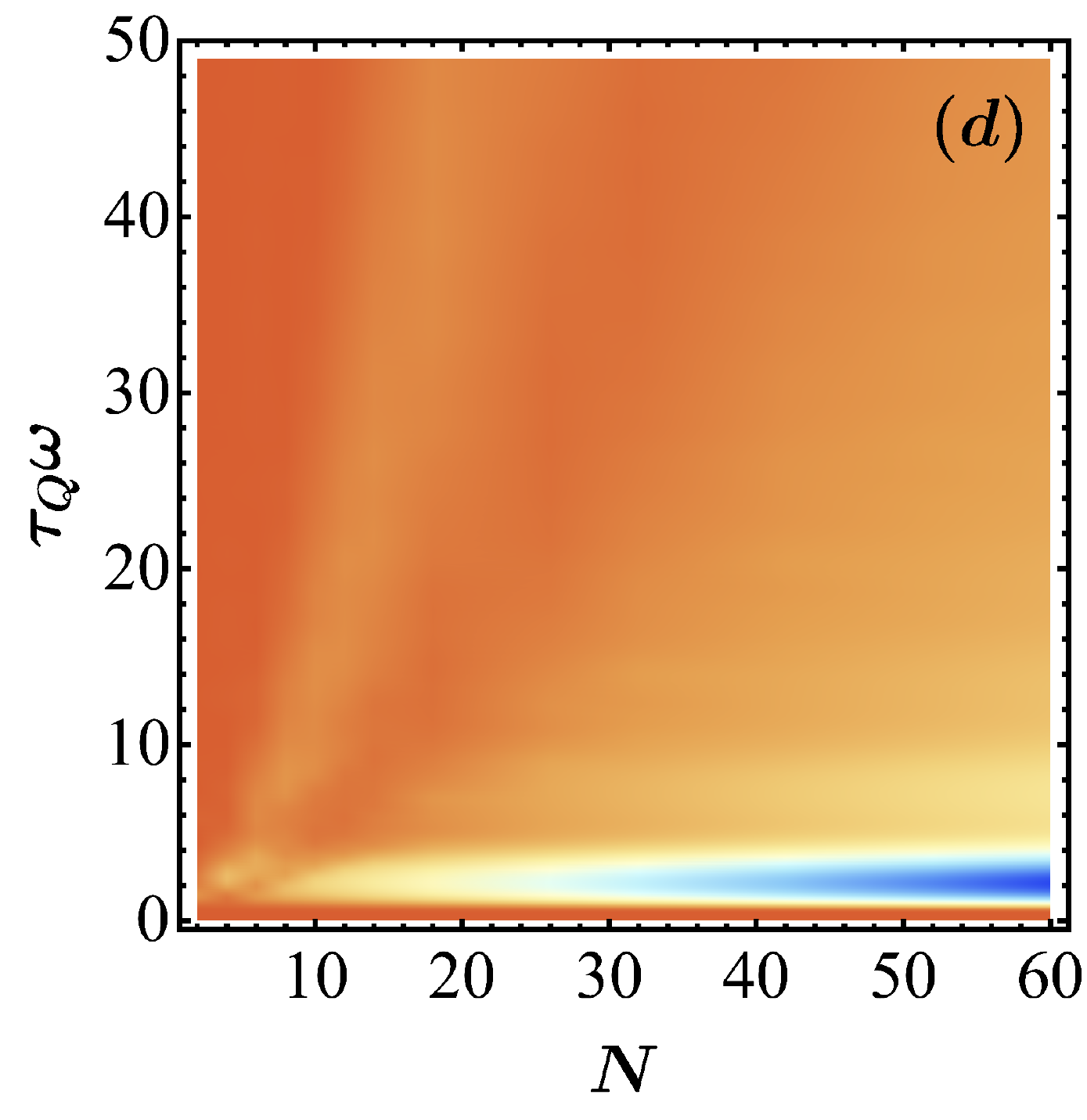}\\
\hspace*{0.15cm}\includegraphics[width=0.45\linewidth]{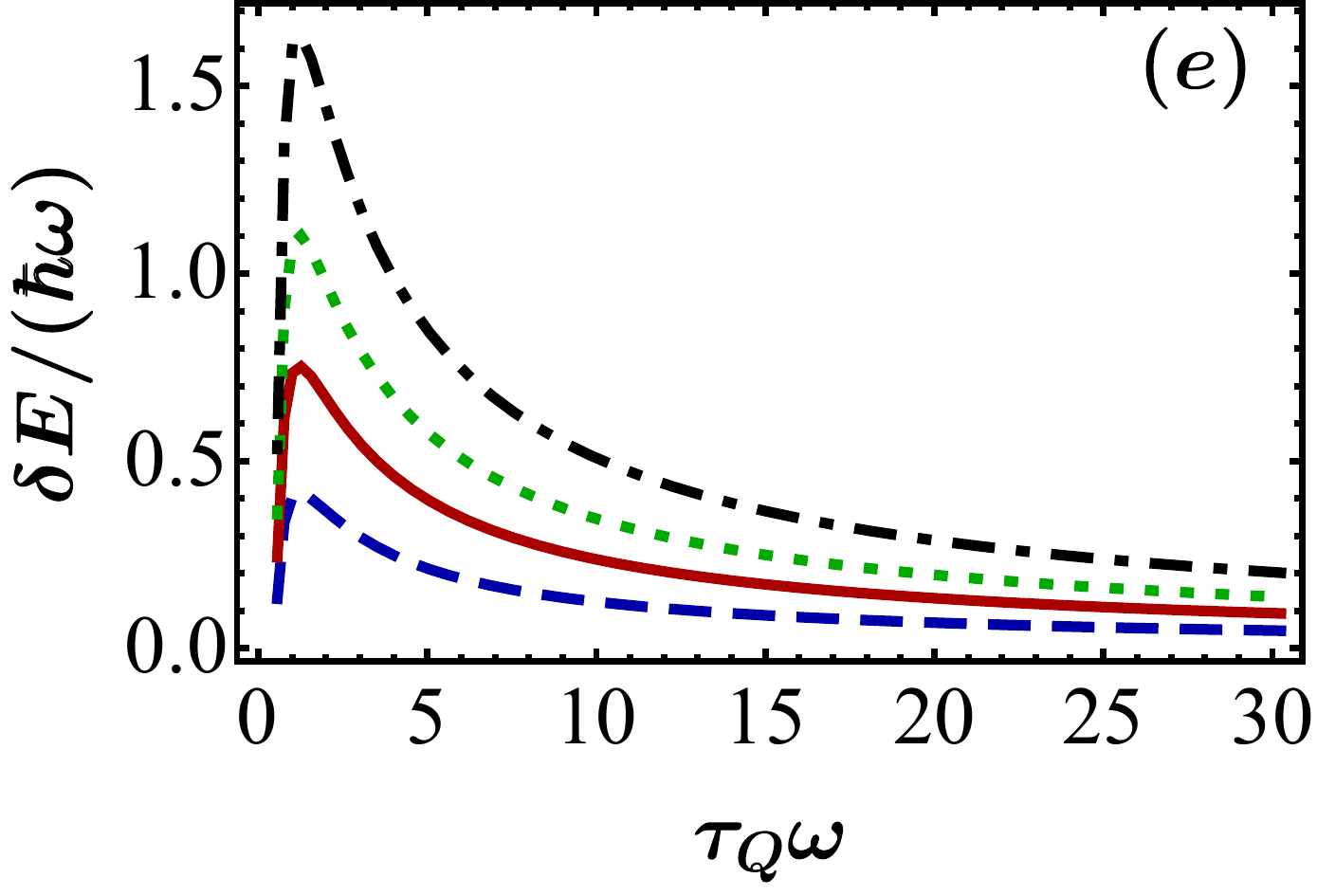}\hspace{0.35cm}\includegraphics[width=0.45\linewidth]{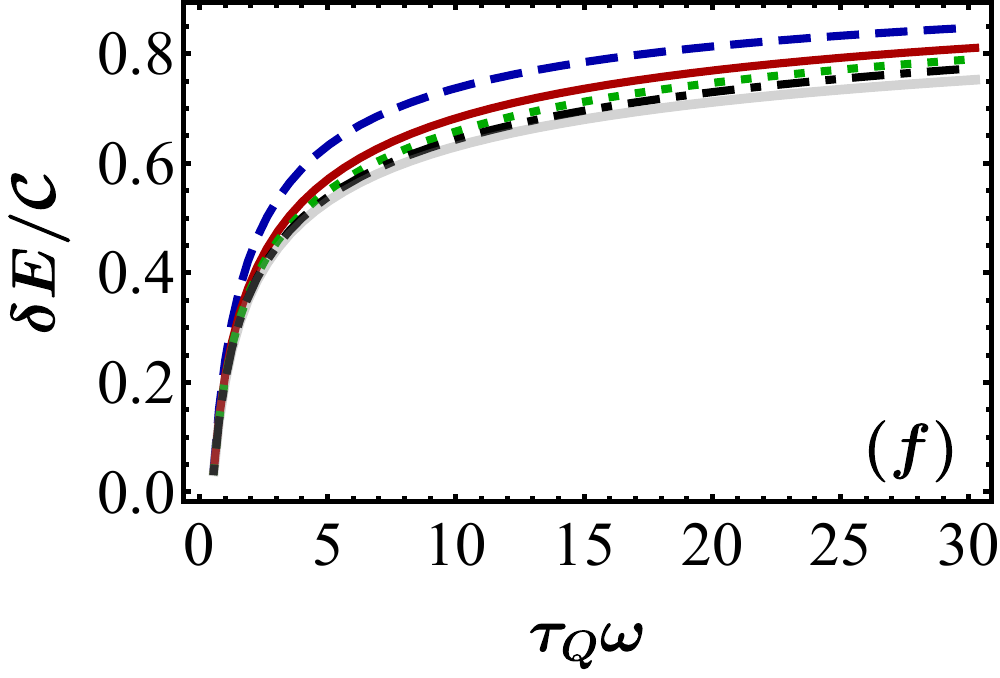}
\end{center}
\caption{The instantaneous fidelity of TFIM for no control (blue dashed line), impulse control  (red solid line) and full control  (green dotted line), for different quench times. Orange shaded area indicates the impulse regime $[t_{-},t_{+}]$ (a) $\omega \tau_Q\!=\!10$ (b) $\omega \tau_Q\!=\!25$ for $N\!=\!16$. Panels (c) and (d) show the final state fidelity for TFIM versus quench time $\tau_Q$ and system size $N$ for the  (c) uncontrolled case and (d) impulse control. Panels (e) and (f): Energetic savings versus quench time $\tau_Q$. $N=4,8,12,18$ (blue dashed line, red solid line, green dotted line, black dot-dashed line) and thermodynamic limit (light gray thick solid line). (e) Savings $\delta E$ (f) relative savings $\delta E / \mathcal{C}$. In all panels, $g_0=0$ and $m=100 \omega^{-1}$}
\label{fig_Ising_density}
\end{figure}

In Fig.~\ref{fig_Ising_density}(a) and (b) we show the fidelity with the instantaneous ground state for the three cases of no control, full control, and impulse control for a system size of $N\!=\!16$, where qualitatively similar behaviors with the LZ model are exhibited. By employing control only during the impulse regime the most detrimental period of defect formation is suppressed and good target state fidelities are achieved.
The effectiveness of impulse control is thoroughly demonstrated by comparing Fig.~\ref{fig_Ising_density}(c) and (d). Here we show the final target state fidelity as a function of system size and quench duration. When no control is applied, i.e. $\kappa\!=\!0$ shown in panel (c), we see that defects rapidly form for larger systems due to the effect of the impulse regime, leading to small final fidelities (lighter, blue region). These results are well described by the Landau-Zener formula applied to the lowest momentum subspace $\mathcal{F}(\tau_Q) \!\approx\! 1- \exp\left[-\frac{2 \pi \omega}{\fabs{\dot{g}}} \sin^2\left(\frac{\pi}{N}\right) \right]$~\cite{DziarmagaPRL}. Employing impulse control provides a significant increase in the final state fidelities, cfr. Fig.~\ref{fig_Ising_density}(d). For extremely short quench times, $\tau_Q\omega\!<\!1$, the impulse regime dominates the dynamics and thus the control term is effectively on for the entire protocol duration. There is then a region of low-fidelity (blue-coloured) for $1\!<\!\tau_Q\omega\!<\!6$ for sufficiently large system sizes. In this region the rapid losses in fidelity during the short adiabatic regimes are too severe to be recovered. Nevertheless, beyond  this small pathological region in parameter space, impulse-only control is highly effective in comparison to uncontrolled evolution, consistently outperforming the uncontrolled case for a range of longer quench times. However, similar to the LZ case, once as approach adiabatic timescales the uncontrolled case can have a slightly higher fidelity than impulse control (upper left quadrant of Fig.~\ref{fig_Ising_density}(c) vs (d)). As previously noted in the LZ setting, this is due to eigenstate population being approximately constant leaving the impulse regime, removing any possibility to recover any lost fidelity from the initial period of free evolution.

We now focus on the energetic costs. In Fig.~\ref{fig_Ising_density} we see that the absolute, panel (e), and relative, panel (f), energetic savings are consistent with the behavior exhibited in the LZ case. We see from Fig.~\ref{fig_Ising_density}(e) that the energetic savings are extensive with the size of the system, however the relative savings exhibits a clear converging, intensive behavior. Nevertheless, a significant saving in the energetic overheads can be achieved while still achieving effective control. Similar to the LZ model, exact expressions for the cost measures in this case can be determined, see Appendix~\ref{Cost_app}.

\begin{figure}[t]
\begin{center}
\includegraphics[width=0.45\linewidth]{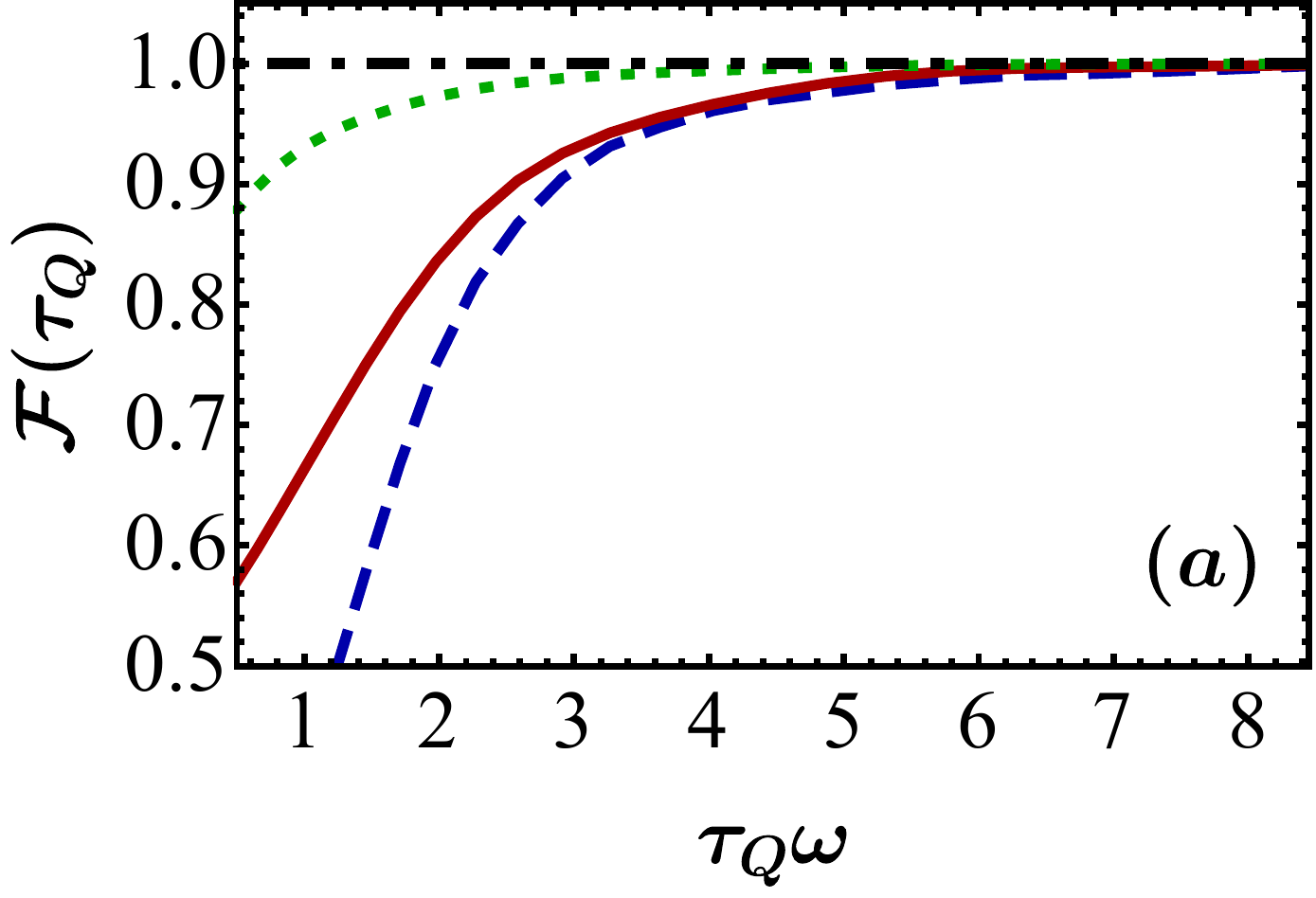}\hspace{0.5cm}\includegraphics[width=0.45\linewidth]{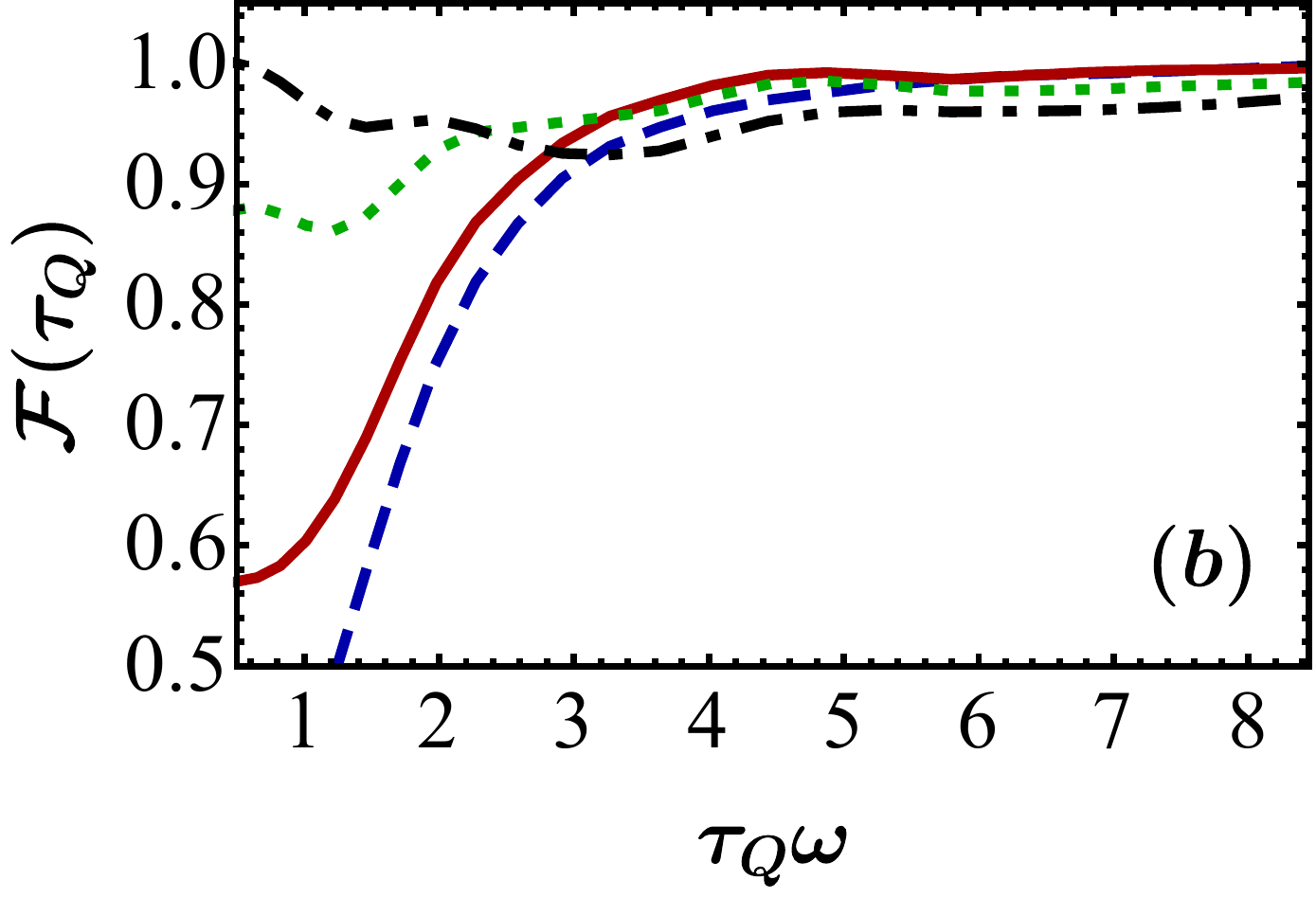}\hspace{0.5cm}\includegraphics[width=0.45\linewidth]{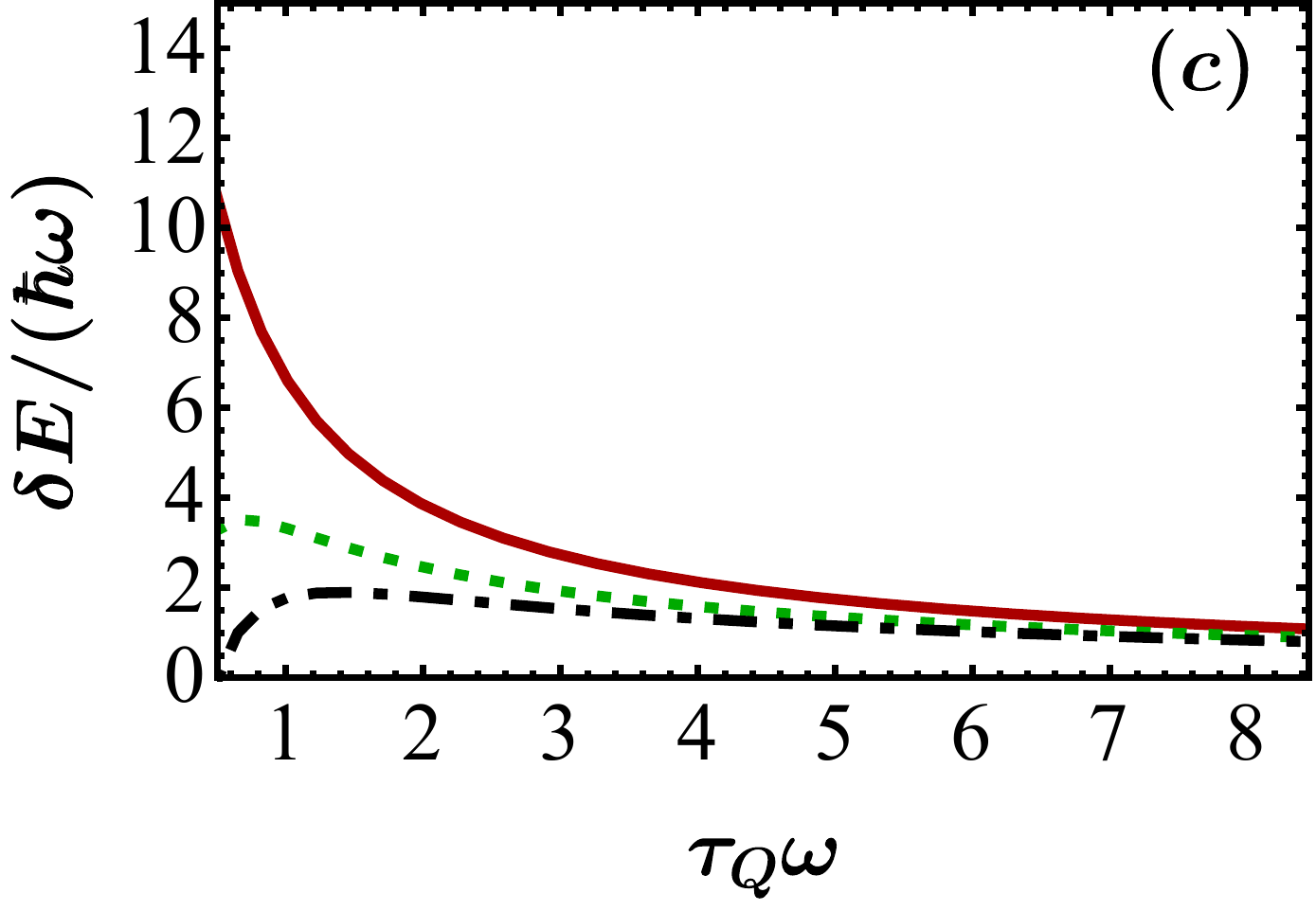}\hspace{0.5cm}\includegraphics[width=0.45\linewidth]{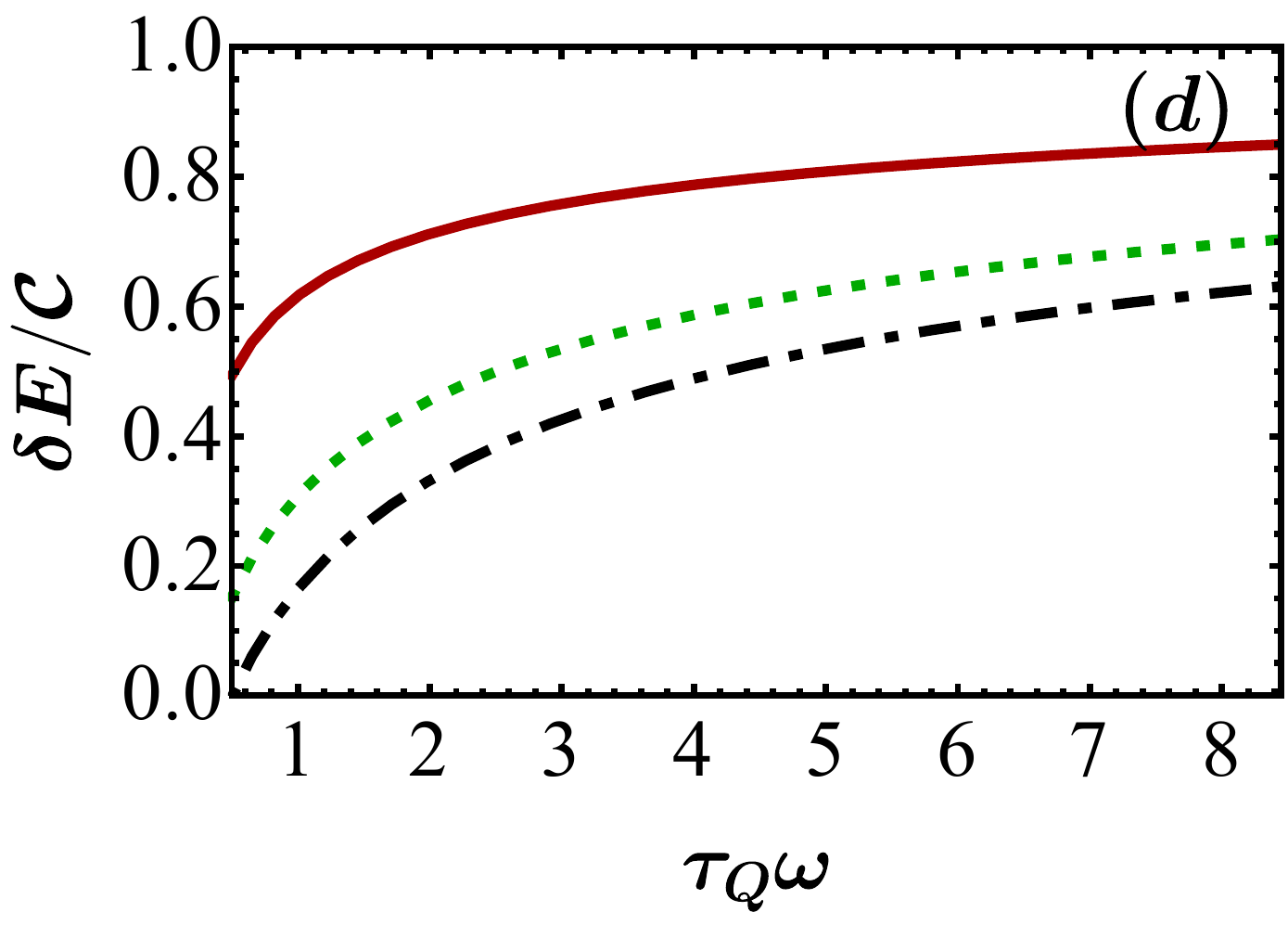} \hspace{0.5cm}
\end{center}
\caption{ The final state fidelity for the $N=6$ TFIM versus quench time for uncontrolled evolution (dashed, blue), $M\!=\!1$ (i.e. two-body control, solid red), $M\!=\!2$ (three body control, dotted green) and $M\!=\!3$ (i.e. full control, dot-dashed black). In panel (a) we show the performance when the control term is always on for the entire evolution, while panel (b) corresponds to impulse control. Panels (c) and (d) show the energetic savings versus quench time $\tau_Q$ for the the same truncated-range impulse control protocol in (b), with the same colour scheme as before. (c) Savings $\delta E$ relative to employing full range, full quench cost $\mathcal{C}$ (d) relative savings $\delta E / \mathcal{C}$. Other parameters: $g_0=0.01$ and $m=100 \omega^{-1}$}
\label{fig_trunc}
\end{figure}

Finally, we investigate the effect of further restricting the counterdiabatic Hamiltonian. By exploiting the form of the counterdiabatic term given by Eqs.~\eqref{eq:trunc0}-\eqref{eq:trunc3}, we can truncate the control terms to restricted range(s) $M$. For clarity, we consider $N=6$ although remark that we expect qualitatively similar behaviors to hold for larger systems. In Fig.~\ref{fig_trunc}(a) we plot the final state fidelity for a range of quench times, employing the control terms for the entire quench. In line with intuition, the fidelities arrange themselves into a hierarchy for short quench times. The uncontrolled case performs the worst, while longer range more complex control works increasingly well until it achieves perfect final fidelities for full control ($M\!=\!3$ in this case), with the relative difference in performance reducing as we approach the the adiabatic limit. For the case of impulse control, Fig.~\ref{fig_trunc}(b),  the same hierarchy holds for very fast protocols. However, as the quench time is increased we see several crossovers in relative performance, indicating that for such intermediate quench times, impulse control exhibits a ``less is more" behavior whereby better (although not perfect) target state fidelities can be achieved by employing a simpler control term in the impulse regime and significant energetic savings can be achieved, cfr. Fig.~\ref{fig_trunc}(c) and (d). 

\section{Conclusion \label{con}}
We have demonstrated that high fidelity coherent control can be achieved at a lower resource overhead by restricting the application of control techniques to when they are strictly necessary. By exploiting the framework provided by the Kibble-Zurek mechanism, which divides the dynamical response of a system driven through a critical point into adiabatic and impulse regimes, we have shown that high target state fidelities can be achieved by only implementing control during the impulse regime. The intuition for this effect relies on the underlying physical principles of the KZM; the adiabatic regime is characterized by a dynamics which is varying sufficiently slow, compared to the energy gap, such that the system is still able to relax. Under these conditions, even though the system may transiently generate some excitations, the system recovers--a remarkably generic feature of adiabatic protocols~\cite{MolmerPRA}. In contrast, control is essential in the impulse regime.  Due to the typically high energetic cost associated with various control protocols~\cite{Campbell2017, Abah2019}, we have shown that significant energetic savings can be achieved using impulse control without significantly sacrificing efficacy. 

\acknowledgements
This work is supported by the Irish Research Council Project ID GOIPG/2020/356, the Science Foundation Ireland Starting Investigator Research Grant ``SpeedDemon" No. 18/SIRG/5508, and the Thomas Preston Scholarship.

\appendix

\section{Cost measure expressions \label{Cost_app}}
In the following, we will present analytical expressions for the energetic cost measures. We assume for simplicity that $S(t)$ is exactly a step function and $g_0<g_c$.

\subsection{Landau-Zener Model}
We begin by noting that the relevant integral can be computed as
\begin{eqnarray}
\int_{t_1}^{t_2} \norm{H_{CD}(s)}ds = \hbar \sqrt{2}  \left\{\arctan\left[\frac{g(t_2)}{\Delta}\right]-\arctan\left[\frac{g(t_1)}{\Delta}\right]\right\}. \nonumber \\
\end{eqnarray}
From this it is clear that the total cost can be written as
\begin{eqnarray}
\mathcal{C}= - \frac{ \sqrt{2} \hbar}{\tau_Q} \arctan\left(\frac{g_0}{\Delta}\right).
\end{eqnarray}
Similarly, the relative savings in this case are
\begin{eqnarray}
\delta E = \frac{\sqrt{2} \hbar}{\tau_Q} \left\{ \arctan\left[\frac{g(t_-)}{\Delta}\right]-\arctan\left(\frac{g_0}{\Delta}\right)\right\}.
\end{eqnarray}
Finally then, the relative savings become
\begin{eqnarray}
\delta E / \mathcal{C} = 1- \frac{\arctan\left[g(t_-)/\Delta\right]}{\arctan\left(g_0/\Delta\right)}.
\end{eqnarray}
These analytic expressions match exactly with the numerical results shown in Fig. \ref{fig_LZM_compare} (b).

\subsection{TFIM}

 Working in the momentum subspace picture the norm can be written as $\norm{H_{CD}}\!=\!\sum_{k>0}\norm{H_{CD,k}}$. The associated energetic cost, Eq.~\eqref{cost}, is then~\cite{Puebla2020}
\begin{equation}
\mathcal{C}=\frac{\hbar}{\sqrt{2}\tau_Q} \sum_{k>0} \int_{0}^{\tau_Q} ds  \fabs{\frac{\dot{g}\sin(kb)}{g(s)^2-2 g(s)\cos(kb)+1}}.
\end{equation}
This can be rewritten as
\begin{eqnarray}
&\mathcal{C}=\frac{\hbar}{\sqrt{2}\tau_Q} \sum_{k>0} \left\{ \arctan\left[\frac{g(\tau_Q)-\cos(k b)}{\sin(k b)}\right] - \arctan\left[\frac{g_0-\cos(k b)}{\sin(k b)}\right]\right\}. \nonumber \\
\end{eqnarray}
The extensive nature of this and the absolute savings, Eq.~\eqref{abs_saving} can be explicitly seen by noting that in the thermodynamic limit we can make the replacement $\sum_{k>0} \rightarrow \frac{N}{2 \pi} \int_0^\pi d(kb)$,
\begin{eqnarray}
\delta E \approx \frac{\hbar N}{2\sqrt{2}\pi\tau_Q}  \left\{
\Phi[g(\tau_Q)]-\Phi[g(t_+)]+\Phi[g(t_-)]-\Phi[g_0]
 \right\}, \nonumber \\
\end{eqnarray}
where we have defined $\Phi[g]=\int_0^\pi dx \arctan\left[\frac{g-\cos(x)}{\sin(x)}\right]$. In this limit then, the relative cost becomes
\begin{eqnarray}
\delta E/\mathcal{C} = 1- \frac{\Phi[g(t_+)]-\Phi[g(t_-)] }{\Phi[g(\tau_Q)]-\Phi[g_0]},
\end{eqnarray}
which is clearly intensive. These expressions agree with the numerical results shown in Fig.~\ref{fig_Ising_density}.\\ 
\section{TFIM technical details \label{TFIM_app}}
The transformations used to determine the counterdiabatic driving term are done by first rotating around the $y$ axis to map $\sigma_i^z \rightarrow \sigma_i^x$ and $\sigma_i^x \rightarrow -\sigma_i^z$ and substituting
\begin{eqnarray}
\sigma_j^x &=& 1-2 c_j^\dagger c_j, \\
\sigma_j^z &=& -(c_j+c_j^\dagger) \prod_{m<n} \left(1-2 c_m^\dagger c_m \right),
\end{eqnarray}
where $c_j^\dagger$ and $c_j$ and are fermionic creation and annihilation operators respectively at site $j$. We then perform a discrete Fourier transformation $c_k=\frac{1}{\sqrt{N}} \sum_j e^{-i k b j} c_j$, where $b$ is the inter-spin spacing and exploit the $\mathds{Z}_2$ parity symmetry. This decouples the Hamiltonian as $H_0=\bigoplus_{k>0} \Psi_k^\dagger H_{0,k} \Psi_k$ where $\Psi_k^\dagger=\left(c_k^\dagger, c_{-k}\right)$. Each momentum subspace is governed by a LZ type Hamiltonian $H_{0,k}=h_k^x\sigma_k^x-h_k^z(g)\sigma_k^z$ where $h_k^z(g)=2\hbar \omega[g-\cos(kb)]$ and $h_k^x=2\hbar\omega\text{sin}(kb)$. Note that the momentum only take on discrete values $k_n=\frac{\pi(2n-1)}{Nb}$ for $n=1, \ldots, N/2$.
This form leads to $H_{CD,k}=\hbar \dot{\theta}_k \sigma_k^y$, Eq.~\eqref{Hcd_Is}, for each momentum subspace analogous to the LZ.

The eigenstates of $H_{0,k}$ are given by
\begin{eqnarray}
 \ket{\phi_{0,k}(t)} &=& \cos\left[\theta_k(t)\right]\ket{0}_k+\sin\left[\theta_k(t)\right] \ket{1}_k, \\
  \ket{\phi_{1,k}(t)} &=& \sin\left[\theta_k(t)\right] \ket{0}_k- \cos\left[\theta_k(t)\right]\ket{1}_k, 
\end{eqnarray}
where $\tan\left[\theta_k(t)\right]=\left[h_k^z-\sqrt{h_k^{x,2}+h_k^{z,2}}\right]/h_k^x$. The ground state of the system is then given by $\ket{\phi_0(t)}=\bigotimes_{k>0} \ket{\phi_{0,k}(t)}$. If the evolved state of the system is written as $\ket{\psi(t)}=\bigotimes_{k>0} \ket{\psi_k(t)}$, then the fidelity becomes
\begin{equation}
\mathcal{F}(t)= \fabs{\braket{\psi(t)}{\phi_0(t)}}^2= \fabs{\prod_{k>0}\braket{\psi_k(t)}{\phi_{0,k}(t)}}^2.
\end{equation}
    
\bibliography{truncated_control.bib}
\end{document}